\documentclass[twocolumn,twocolappendix]{aastex63}  % linenumbers

\usepackage{amssymb,amsmath,bm}
\usepackage{graphicx}
\usepackage{natbib}
\usepackage{color}
\usepackage{soul}

\usepackage{algorithm}
\usepackage[noend]{algpseudocode}
\usepackage{tikz}
\usetikzlibrary{quantikz}
\usepackage{savesym}
\savesymbol{splitbox}
\usepackage{adjustbox}
\restoresymbol{TXF}{splitbox}
\usepackage{hyperref}
\def\aap{A\&A}

\def\apjs{ApJS}
\def\apjl{ApJL}

\received{xxx xx, 2026}
\revised{xxx xx, 2026}
\accepted{xxx xx, 2026}
\submitjournal{ApJ}

\begin{document}

% SWITCH THESE ON TO INCLUDE REFEREE RESPONSE AT FRONT OF ARTICLE
%%%%%%%%%%%%%%%%%%%%%%%%%%%%%%%%%%%%%%%%%%%%%%%%%%%%%%%%%%%%%%%%%%%%%%%%%%%%%%%%%%%%%%%%      
%
%{\onecolumngrid \color{purple}   \input{refResponseText.tex}}
%
%%%%%%%%%%%%%%%%%%%%%%%%%%%%%%%%%%%%%%%%%%%%%%%%%%%%%%%%%%%%%%%%%%%%%%%%%%%%%%%%%%%%%%%%      
%
%   TO GENERATE GENERATE REFEREE RESPONSE, DOWNLOAD THE PROJECT AND DO
%   latexdiff letterVersion2019-08-28.tex letter.tex > refResponse2.tex
%   and then compile refResponse2.tex with LaTeX
%
%%%%%%%%%%%%%%%%%%%%%%%%%%%%%%%%%%%%%%%%%%%%%%%%%%%%%%%%%%%%%%%%%%%%%%%%%%%%%%%%%%%%%%%%      

%##############################################################################

\title{Cosmo-SPINN: Fuzzy Dark Matter Simulations with Physics-Informed Generative Networks}

\correspondingauthor{Ashutosh K. Mishra}
\email{ashutosh.mishra@epfl.ch}

\author[0009-0002-8819-8236]{Ashutosh K. Mishra}
\affiliation{Institute of Physics, Laboratory of Astrophysics, École Polytechnique Fédérale de Lausanne (EPFL), Observatoire de Sauverny, Versoix, 1290, Switzerland}

\author[0000-0002-1027-1213]{Emma Tolley}
\affiliation{Institute of Physics, Laboratory of Astrophysics, École Polytechnique Fédérale de Lausanne (EPFL), Observatoire de Sauverny, Versoix, 1290, Switzerland}
% \affiliation{SCITAS, École
%  Polytechnique Fédérale de Lausanne (EPFL) , Lausanne, 1015, Switzerland}
\author[0009-0004-1864-512X]{Nicolas Cerardi}
\affiliation{Institute of Physics, Laboratory of Astrophysics, École Polytechnique Fédérale de Lausanne (EPFL), Observatoire de Sauverny, Versoix, 1290, Switzerland}
% \affiliation{SCITAS, École
%  Polytechnique Fédérale de Lausanne (EPFL) , Lausanne, 1015, Switzerland}
%##############################################################################

\begin{abstract}
Generative machine learning models have recently emerged as powerful tools for producing cosmological simulations. However, many existing emulators do not explicitly enforce the underlying physical dynamics governing cosmological evolution, often leading to artifacts and poor adherence to the evolution equations. In this work, we present a physics-informed generative U-Net framework for fuzzy dark matter (FDM) that addresses two complementary tasks: (i) the evolution of cosmological fields from initial conditions to an arbitrary cosmological scale factor and (ii) the super-resolution of FDM simulations at a specified cosmological scale factor. Our model incorporates a physics-informed loss function that explicitly enforces consistency with the underlying Schrödinger-Poisson (SP) dynamics during training. For the evolution task, we find that the inclusion of this physics-based loss significantly improves the quality of the predicted simulations, even when only a small amount of training data is available. Using only 20\% of the training data, the model accurately reproduces the target simulations in a 1 $h^{-1}$Mpc box. Furthermore, the framework generalizes effectively across previously unseen realizations of the initial conditions. For the super-resolution task, we present, for the first time, a generative super-resolution model trained on FDM simulations obtained by solving the full SP equations, considering both single and multiple realizations of the initial conditions and analyzing the role of the physics-informed loss in each case. Our approach enables modern generative modeling of cosmological simulations while maintaining physical consistency and substantially reducing generative artifacts.

\end{abstract}
\section{Introduction}\label{intro}
The nature of dark matter (DM) remains one of the central unsolved problems in cosmology. Beyond the standard Cold Dark Matter (CDM) paradigm, numerous alternative DM candidates have been proposed (see, e.g., \citet{2024_Cirelli} for a review). Among the many proposed candidates, ultralight scalar particles such as axion-like fields have attracted significant attention across a wide range of masses \citep{2010_Axiverse,2021_Ferreira, 2021_Hui,2025_Eberhardt}. A particularly well-studied type is ultralight fuzzy dark matter (FDM), corresponding to particle masses in the range $10^{-22}eV \lesssim m \lesssim 10^{-20} eV$, where quantum effects become astrophysically relevant and give rise to wave-like behavior on galactic scales. In the non-relativistic regime, the dynamics of this ultralight dark matter are governed by the Schr\"odinger--Poisson (SP) equations.  FDM retains the large-scale structure of CDM while modifying small-scale structure formation, such as producing solitonic halo cores \citep{2014_Schive,2017_Hui}, potentially alleviating several small-scale tensions of the standard cosmological model. 

These characteristic predictions have made FDM an attractive alternative to CDM. While FDM has been shown to successfully reproduce the velocity dispersion profiles of the Milky Way \citep{DEMARTINO2020100503} and a number of dwarf galaxies \citep{Chen_2017}, recent observational analyses have placed increasingly stringent constraints on ultralight dark matter models. Using stellar kinematics of ultra-faint dwarf galaxies together with full SP simulations, \citet{2025_MayBounds} derived lower bounds on the dark matter particle mass as strong as $m > 8 \times 10^{-18} $ eV under certain assumptions, significantly constraining the canonical FDM regime. The systems used in this analysis were the dwarf galaxies Segue~1 and Segue~2, together with the ``micro-galaxy'' Ursa Major~III/UNIONS~1. However, the interpretation of these constraints remains uncertain, as both Segue~1 and Ursa Major~III may not be dark-matter-dominated systems\citep{Lujan_2025,2022_Goldstein}, contrary to the assumptions adopted by \citet{2025_MayBounds}. Consequently, the stringent bound of $m > 8 \times 10^{-18} \ \mathrm{eV}$ should be regarded as provisional and subject to validation with future observations. Nevertheless, heavier regions ($m>10^{-18}$ eV) of the broader wave dark matter parameter space  remain of considerable interest, as does FDM itself in scenarios involving mixed dark matter models, partial ultralight dark matter fractions, or alternative scalar-field realizations \citep{2026_Crumrine,2026_Johnston}. Exploring these possibilities requires efficient and physically reliable SP simulation techniques across a wide range of initial conditions and resolutions.

Over the past decade, numerical simulations based on the SP equations have become indispensable for understanding nonlinear structure formation in FDM models \citep{2014_Schive, Mocz_2017_BECDM,2021_May, 2026_Schive}. However, these wave-based simulations remain computationally expensive because the de Broglie wavelength must be resolved throughout the simulation volume. Consequently, cosmological SP simulations are currently limited to relatively small box sizes compared to conventional CDM N-body simulations. Although hybrid approaches combining wave and fluid formulations have recently improved computational efficiency \citep{2024_Kunkel}, achieving large-volume, high-resolution SP simulations remains challenging. As a result, there are far fewer high-resolution FDM simulations than their CDM counterparts, significantly limiting studies of small-scale structure formation and halo properties in wave dark matter models.

Deep learning techniques have shown considerable promise in accelerating cosmological simulations and emulation tasks \citep{doi:10.1073/pnas.1821458116,dai2020,necola}. 
Super-resolution (SR) techniques using machine learning offer a promising prospect of overcoming the computational limitations of FDM numerical simulations. Originally developed for image processing, super-resolution methods use neural networks to reconstruct high-resolution (HR) features from low-resolution (LR) inputs. Recently, these methods have been adapted for cosmological simulations using deep generative models, particularly generative adversarial networks (GANs) conditioned on low-resolution simulations \citep{2020_Kodi,2021_Li,2023_Sipp}. Such approaches have successfully reproduced key statistical properties of CDM simulations, including the matter power spectrum and halo mass functions, while achieving speed-ups of several orders of magnitude relative to full high-resolution N-body simulations. However, these approaches have so far been explored primarily in the context of CDM, where the underlying dynamics are comparatively simpler and do not involve wave interference phenomena. Only limited work has considered FDM. For example, \citet{2023_Sipp} developed a GAN for FDM super-resolution, but without training their model on FDM simulations that incorporate the full quantum pressure. In addition, the SR mapping remains an inherently ill-posed problem, as multiple high-resolution realizations can correspond to the same low-resolution simulation after downsampling. This ambiguity is further worsened for FDM when learning is performed solely in density space, since density is agnostic to phase of the FDM fields involved. Consequently, effective super-resolution for FDM requires learning strategies that exploit information beyond the density field alone.

Furthermore, standard neural networks often fail to preserve the underlying physical dynamics, leading to unphysical artifacts and poor generalization \citep{LINKA_2022, 2022_Faroughi}. Physics-informed machine learning addresses these limitations by incorporating governing differential equations directly into the learning process. In our previous work, SPINN \citep{2025_SPINN}, we demonstrated for the first time that physics-informed neural networks (PINNs) can successfully model gravitational collapse governed by the SP equations and reproduce key features of wave dark matter dynamics. That work established the viability of physics-informed approaches for cosmological SP simulations. 

Building on these developments, we extend these ideas toward generative cosmological emulation. In this work we present Cosmo-SPINN, a Schrödinger--Poisson informed generative framework for FDM simulations based on a U-Net architecture with a physics-informed loss function. The framework is applied to two tasks: (i) the evolution of cosmological fields from initial conditions to a specified scale factor and (ii) the super-resolution of FDM simulations. Unlike conventional emulators trained purely on data, our framework explicitly enforces SP dynamics during training, substantially reducing generative artifacts while improving data efficiency and generalization across realizations of the initial conditions. Furthermore, we demonstrate, for the first time, super-resolution FDM simulations generated from full SP evolutions using a physics-informed generative model. Our study focuses on redshifts within the Epoch of Reionization, where the suppression of small-scale structure by FDM can leave observable signatures \citep{2021_Jones}, making efficient emulators particularly relevant for constraining FDM models. Our results show that physically informed generative approaches provide a promising pathway toward accelerated large-scale SP simulations. Our implementation leverages established machine learning libraries, notably PyTorch \citep{2019_Paszke}, enabling more efficient and accurate solutions.

This paper is structured as follows. In Section \ref{Bkgd}, we present the governing equations for the FDM used in this work, provide an overview of the numerical scheme used to solve them and the framework of Physics Informed Generative Networks. In Section \ref{method}, we describe the initial conditions, the SP simulations used for training and testing, and the Cosmo-SPINN framework, including the physics-informed Evolution and Super-Resolution models, their architectures, and optimization procedures. We present the results and discussion in Sections \ref{Res} and  \ref{Disc} respectively before concluding in Section \ref{conc}. Additional validation tests for the Evolution Model, including performance across redshifts (Appendix \ref{appendix_A}), and generalization to unseen realizations of the initial conditions (Appendix \ref{appendix_B}), are provided in the appendices.

\section{Theoretical Background}\label{Bkgd}
\subsection{Governing Equations}\label{GE}
Fuzzy DM consists of an ultralight boson with  mass $m \ll 30$ eV resulting in a macroscopic de Broglie wavelength. An axion-like particle is one possible candidate for FDM. The axion $\phi$ is a real angular field with a periodicity of f, f being an axion-decay constant. Taking a primordial value of order f, the axion transitions from being frozen early in the universe to oscillating in the late times, resulting in a relic density of \citep{2010_Axiverse,2015_Marsh,2021_Hui}:

\begin{align}
    \Omega_{\text{axion}} \sim 0.1 \left(\frac{f}{10^{17}\text{GeV}}\right)^2 \left(\frac{m}{10^{-22}\text{eV}}\right)^{1/2}
\end{align}

Assuming non-relativisitic regime and t as the proper time, we introduce complex-valued scalar field or wavefunction
 $\psi(\boldsymbol{x},t)  \in \mathbb{C}$ with $\boldsymbol{x}\in \mathbb{R}^{3}$ that relates to the real axion field $\phi$ as:
\begin{align}
    \phi = \sqrt{\frac{\hbar^3 c}{2m}} \left(\psi e^{-i\frac{mc^2}{\hbar}t} + \psi^{*} e^{i\frac{mc^2}{\hbar}t}\right)
\end{align}
where, $\hbar$ is the reduced Planck's constant and $c$ is the speed of light in vacuum. Under the condition $|\dot{\psi}| \ll m |\psi|$, and adopting the perturbed Friedmann-Lemaître-Robertson-Walker (FLRW) metric,
\begin{align}
    \textrm{d}s^2 = \left(1+\frac{2V}{c^2}\right)c^2 \textrm{d}t^2 - a(t)^2 \left(1-\frac{2V}{c^2}\right)\textrm{d}\boldsymbol{x}^2
\end{align}
the Klein-Gordon equation for the axion $\phi$ gives the following Schr\"odinger-Poisson equations:
\begin{align}\label{SE}
    i \hbar \partial_t \psi(\boldsymbol{x},t) &= -\frac{\hbar^2}{2ma^2}\nabla^2\psi(\boldsymbol{x},t) + \frac{m}{a}V\psi(\boldsymbol{x},t) 
\end{align}
\begin{align}\label{PE}
        \nabla^2 V(\boldsymbol{x},t) &= 4\pi G m (|\psi(\boldsymbol{x},t)|^2 - \langle|\psi|^2\rangle(t))
\end{align}
where $a$ is the cosmological scale factor, $V$ is the Newtonian gravitational potential,
the angle brackets in $\langle|\psi|^2\rangle$ indicate the spatial average, and in the above equations, all the quantities and coordinates are in the ``comoving" form which are related to ``physical" quantities as:
$$
\boldsymbol{x} = a^{-1}\boldsymbol{x}_{\text{phys}}, \, \nabla = a\nabla_{\text{phys}}, \, \psi = a^{3/2}\psi_{\text{phys}}, \, V = aV_{\text{phys}}
$$
Despite the appearance of Schr\"odinger equation, the wavefunction $\psi$ should be interpreted as a classical field. $\psi$ is subject to wave effects such as interference, similar to interference of waves in classical electromagnetism. It should be noted that the Eqs. (\ref{SE} \& \ref{PE}) describes the evolution of a single macroscopic wavefunction of a Bose-Einstein condensate, with a mass density $\rho = m|\psi|^2 $,   rather than the wavefunction of an individual particle. One can also introduce super-comoving time $d\tilde{t} = a^{-2} dt$ \citep{1998_Martel} and $\tilde{V} = aV$, to obtain the more familiar Schr\"odinger equation without scale factors:
\begin{equation}\label{SSE}
    i \frac{\partial\psi}{\partial \tilde{t}} = \left(-\frac{\hbar^2}{2m}\nabla^2 +\frac{m}{\hbar}\tilde{V}\right)\psi
\end{equation}
This formulation can be useful because it removes scale factors in the denominator which makes controlling error at low redshifts very difficult. 

The Schrödinger equation implies a conservation law described by the continuity
equation
\begin{equation}\label{CE}
\partial_t \rho + \nabla \cdot (\rho \mathbf{v}) = 0,
\end{equation}
with the density current
\begin{equation}\label{DC}
\rho \mathbf{v} =
\frac{\hbar}{2 i}
\left(
\psi^* \nabla \psi
-
\psi \nabla \psi^*
\right),
\qquad
(\rho = m |\psi|^2).
\end{equation}
The (comoving) velocity field $\mathbf{v}$ dictates the peculiar
velocity of matter at each point. Using the \citet{1927_Madelung} transformation:
\begin{equation}\label{MT}
\psi =
\sqrt{\frac{\rho}{m}}\, e^{iS},
\end{equation}
with absolute value $\sqrt{\rho/m}$ and phase S, where
\[
\rho = a^{3} \rho_{\text{phys}} =m |\psi|^2
\]
is indeed the same mass density as in Eq.~\eqref{CE}. Inserting this into the
expression for the current in Eq.~\eqref{DC} yields
\begin{equation}\label{vs}
\mathbf{v} =
\frac{\hbar}{m}\nabla S,
\end{equation}
that is, the gradient of the wave
function's phase determines the velocity field. The wave function written using the Madelung transformation (Eq.~\ref{MT})
, can be used to rewrite the SP Equations yielding the continuity Eq.~\eqref{CE} along with a modified Euler equation:
\begin{equation}
\partial_t \mathbf{v}+\frac{1}{a^2}(\mathbf{v} \cdot \nabla)\mathbf{v}
=-\nabla \Phi+\frac{\hbar^2}{2m^2 a^2}\nabla
\left(\frac{\nabla^2 \sqrt{\rho}}{\sqrt{\rho}}\right).
\end{equation}
(cf. e.g. \citet{Mocz_2017_BECDM}). This allows for a hydrodynamical
interpretation of the density and velocity fields $\rho$ and
$\mathbf{v}$. When the quantum pressure term vanishes, which happens on large scales \citep{2002_Bernardeau,2021_Hui}, the Madelung equations recover the pressureless Euler–Poisson equations, similar to the Vlasov-Poisson equations governing CDM in the limit $\hbar/m \to 0$ and before shell crossing. Despite the fact that the Madelung equations are highly non-linear, they become tractable with existing hydrodynamic codes. However, they are inherently ill-behaved due to the presence of quantum pressure term which blows up as $\rho \to 0$, which is common in the DM overdensity evolution where the density vanishes, especially in the voids between halos and filaments. Therefore we choose to evolve the SP-system of equations (\ref{SE} \& \ref{PE}) instead of the Madelung formalism in our work, in particular with the super-comoving time formalism as it allows us to control the errors better at lower redshifts, permitting the long-time evolution.

It should be noted that Eqs.(~\ref{SE} \& \ref{PE} ) only have a single parameter
given by the constant $\hbar/m$, which is related to the de Broglie
wavelength as
\begin{equation}
\lambda_{\mathrm{dB}} =
\frac{2\pi \hbar}{m \text{v}}.
\end{equation}

Another important scale set by $\hbar/m$ is the FDM Jeans length. In FDM, quantum pressure arising from the Heisenberg uncertainty principle counteracts gravitational collapse below a characteristic scale. As a result, at least to linear order the density perturbations larger than the Jeans length can grow gravitationally, while smaller perturbations remain oscillatory and do not collapse. The corresponding comoving Jeans wavenumber $k_J$ at redshift $z$ (correspondingly at scale factor $a$) is given by \citep{Wayne_2000,2019_Li}:
\begin{equation}\label{jeans_wn}
k_{\mathrm{J}} \equiv
\frac{44.7}{\mathrm{Mpc}}
\left( 6a\frac{\Omega_{m0}}{0.3} \right)^{1/4}
\left( \frac{H_0}{70\,\mathrm{km\,s^{-1}}\textrm{Mpc}^{-1}} \frac{m}{10^{-22}\mathrm{eV}} \right)^{1/2}.
\end{equation}

%Equations applicable to CDM 

\subsection{Numerical Scheme for Simulations}

The simulations are performed within a standard cosmological
$\Lambda$CDM background, where CDM is replaced by FDM inside a periodic cubic domain of side length
$L$. The simulation volume samples the large-scale matter distribution
of the Universe and is initialized with the mean comoving matter
density
\begin{equation}
\langle \rho \rangle
=
\Omega_m \rho_{\mathrm{crit}}
=
\Omega_m
\frac{3H_0^2}{8\pi G}.
\tag{16}
\end{equation}

To evolve the system, we solve the Schrödinger--Poisson (SP)
equations using the second-order unitary split-step pseudo-spectral
method described in \citet{Mocz_2017_BECDM}. The method combines
operator splitting with Fast Fourier Transforms (FFT), allowing the
kinetic and gravitational terms to be evolved independently while
preserving unitary time evolution.

The formal evolution of the wavefunction over a timestep $\Delta t$
is given by
\begin{align}
\psi(t+\Delta t,\mathbf{x})
&=
\mathcal{T}
\exp
\Bigg[
-i
\int_t^{t+\Delta t}
\Bigg(
-\frac{\hbar}{2m}
\frac{1}{a(t')^2}\nabla^2
\nonumber \\[4pt]
&\qquad\qquad\qquad
+
\frac{m}{\hbar}
\frac{1}{a(t')}
V(t',\mathbf{x})
\Bigg)
dt'
\Bigg]
\psi(t,\mathbf{x}).
\end{align}

where $\mathcal{T}$ denotes the time-ordering operator. Using the
Baker--Campbell--Hausdorff expansion, the evolution operator is
approximated to second-order accuracy by separating the kinetic and
potential contributions,
\begin{align}
\psi(t+\Delta t,\mathbf{x})
\approx\;&
e^{-i \frac{m}{\hbar}
\frac{\Delta t}{2a(t)}
V(t+\Delta t,\mathbf{x})}
\nonumber \\[4pt]
&\times
e^{\,i \frac{\hbar}{2m}
\frac{\Delta t}{a(t)^2}\nabla^2}
\nonumber \\[4pt]
&\times
e^{-i \frac{m}{\hbar}
\frac{\Delta t}{2a(t)}
V(t,\mathbf{x})}
\psi(t,\mathbf{x}).
\end{align}
This decomposition corresponds to the standard
``kick--drift--kick'' scheme,
\[
\underbrace{
e^{-i \frac{m}{\hbar}
\frac{\Delta t}{2a(t)}
V(t+\Delta t,\mathbf{x})}
}_{\mathrm{kick}}
\quad
\underbrace{
e^{\,i \frac{\hbar}{2m}
\frac{\Delta t}{a(t)^2}\nabla^2}
}_{\mathrm{drift}}
\quad
\underbrace{
e^{-i \frac{m}{\hbar}
\frac{\Delta t}{2a(t)}
V(t,\mathbf{x})}
}_{\mathrm{kick}}.
\tag{19}
\]

The potential ``kick'' updates are performed in configuration space,
while the kinetic ``drift'' step is evaluated in Fourier space using
FFT methods. Spatial derivatives are therefore computed spectrally,
which improves numerical accuracy and computational efficiency.

At each timestep, the gravitational potential is obtained by solving
the Poisson equation in Fourier space,
\begin{equation}
V
=
\mathrm{IFFT}
\left[
-\frac{1}{k^2}
\,
\mathrm{FFT}
\left(
4\pi G m
\left(
|\psi|^2 - \langle|\psi|^2\rangle
\right)
\right)
\right].
\end{equation}

The numerical evolution over one timestep proceeds as follows:
(i) compute the gravitational potential from the density field,
(ii) apply a half-step potential update (kick),
(iii) evolve the kinetic term for a full timestep in Fourier space
(drift), (iv) recompute the gravitational potential, and
(v) apply a second half-step potential update.
The complete algorithm is summarized below.

\begin{algorithm}[H]
\caption{Second-Order Spectral Method for the Schrödinger--Poisson Equations}
\begin{algorithmic}[1]
\State Initialize wavefunction $\psi$ and compute initial potential V
\While{$t < t_{\mathrm{final}}$}

    \State Compute gravitational potential:
    \[
    V =
    \mathrm{IFFT}
    \left[
    -\frac{1}{k^2}
    \,
    \mathrm{FFT}
    \left(
    4\pi G m
    (|\psi|^2 - \langle|\psi|^2\rangle)
    \right)
    \right]
    \]

    \State Apply half-step potential update (kick):
    \[
    \psi
    \gets
    \exp
    \left[
    -i
    \left(
    \frac{m}{\hbar}
    \right)
    \left(
    \frac{\Delta t}{2a}V
    \right)
    \right]
    \psi
    \]

    \State Apply full-step kinetic evolution (drift):
    \[
    \psi
    \gets
    \mathrm{IFFT}
    \left[
    \exp
    \left(
    -i
    \frac{\hbar k^2}{2ma^2}
    \Delta t
    \right)
    \,
    \mathrm{FFT}(\psi)
    \right]
    \]

    \State Recompute gravitational potential:
    \[
    V =
    \mathrm{IFFT}
    \left[
    -\frac{1}{k^2}
    \,
    \mathrm{FFT}
    \left(
    4\pi G m
    (|\psi|^2 - \langle|\psi|^2\rangle)
    \right)
    \right]
    \]

    \State Apply second half-step potential update (kick):
    \[
    \psi
    \gets
    \exp
    \left[
    -i
    \left(
    \frac{m}{\hbar}
    \right)
    \left(
    \frac{\Delta t}{2a}V
    \right)
    \right]
    \psi
    \]

    \State Advance time:
    \[
    t \gets t + \Delta t
    \]

\EndWhile
\end{algorithmic}
\end{algorithm}

This numerical scheme preserves second-order accuracy and unitary
time evolution. The pseudo-spectral formulation efficiently computes
spatial derivatives and gravitational interactions, making it well
suited for simulations of wave-like dark matter and Bose--Einstein
condensate systems.
The kick operator acts locally in position space and only modifies
the phase of the wavefunction. To avoid phase aliasing, the phase
change per timestep is required to remain below $2\pi$, giving
\begin{equation}\label{tk}
\Delta t_{\mathrm{kick}}
<
C_{\mathrm{kick}}
\frac{2\pi\hbar}{m}
\frac{a}{|V|_{\max}},
\end{equation}
where $|V|_{\max}$ is the maximum potential amplitude and
$C_{\mathrm{kick}} \leq 1$ is a numerical safety factor.

The drift operator is evaluated in Fourier space and contributes a
phase proportional to $k^2$. Requiring the maximum phase change
to remain below $2\pi$ gives
\begin{equation}\label{td}
\Delta t_{\mathrm{drift}}
<
C_{\mathrm{drift}}
\frac{4m}{\pi\hbar}
a^2 \Delta x^2,
\tag{34}
\end{equation}
where $\Delta x$ is the grid spacing and
$k_{\max} = \pi/\Delta x$ is the Nyquist wavenumber. The factor
$C_{\mathrm{drift}} \leq 1$ is an empirical stability coefficient whose value
depends on the implementation scheme and the specific application at hand.

The scaling $\Delta t \propto \Delta x^2$ is characteristic of
diffusion-type equations and corresponds to requiring that the
maximum resolved velocity does not propagate more than one grid
cell per timestep. This is the main reason why generating large scale FDM simulations are computationally extremely expensive and time consuming.
\subsection{Physics Informed Generative Networks}

Physics Informed Generative Networks (PIGNs) extend physics-informed machine learning frameworks by combining generative neural architectures with physical constraints derived from the governing equations of the system. Unlike traditional Physics Informed Neural Networks (PINNs), which directly learn the solution of differential equations through collocation points in space-time, PIGNs are designed to generate physically consistent realizations of complex fields while preserving the statistical and dynamical properties imposed by the underlying physics. Such approaches are particularly useful for high-dimensional cosmological simulations and super-resolution tasks, where reconstructing high-resolution fields from low-resolution inputs is an inherently ill-posed inverse problem. Physics-informed generative models address this ambiguity by generating realizations that satisfy the governing equations while reproducing the correct physical structures and statistical distributions.

In this work, the generative networks are trained to learn a physics-informed mapping between cosmological field realizations while being constrained by the governing SP equations. Depending on the task, the architecture is adapted to either predict the cosmological fields at a specified cosmological scale factor from the initial conditions (evolution model) or reconstruct high-resolution realizations from low-resolution inputs (super-resolution model). In both cases, the networks take as input the relevant field variables together with the scale factor and predict physically consistent realizations of the wavefunction and gravitational potential fields across the simulation domain. Physical information is incorporated through additional loss terms constructed from the residuals of the governing equations, thereby regularizing the learning process and promoting consistency with the underlying dynamics.

The general form of the governing differential equations can be written as
\begin{align}
    \mathcal{D}[G(X,\theta);\Lambda] = f(X), \quad X \in \Omega,
\end{align}
with associated boundary and initial conditions,
\begin{align}
    \mathcal{B}[G(X,\theta)] = g(X), \quad X \in \partial \Omega,
\end{align}
where $G(X,\theta)$ denotes the generative neural network parameterized by $\theta$, $\mathcal{D}$ represents the non-linear differential operator governing the dynamics, $\Lambda$ denotes the physical parameters entering the equations, $f(X)$ is the source term, and $\mathcal{B}$ specifies the boundary or initial condition operator and $g(X)$ specifies the corresponding prescribed boundary or initial condition over the computational domain $\Omega \subset \mathbb{R}^{d+1}$ with boundary $\partial\Omega$.
In our case, the residuals for the SP equations would be:
\begin{align}
    i \left[\frac{\partial}{\partial \tilde{t}} +\frac{\hbar}{2m}\nabla^2 -\frac{m}{\hbar}\tilde{V}\right]\psi &= 0 = f_1(X) \\
            \nabla^2 V(\boldsymbol{x},t) - 4\pi G m (|\psi(\boldsymbol{x},t)|^2 - \langle|\psi|^2\rangle(t)) &=0 = f_2(X)
\end{align}
where $\psi$ is the complex wavefunction and $V$ is the gravitational potential. Periodic boundary conditions are expected to be learned purely from the data loss by the network.

The training objective of the PIGN combines the generative reconstruction loss with physics-based regularization terms. In addition to matching the target high-resolution realizations, the network is penalized for violating the SP equations and the imposed additional physical consistencies such as mass conservation and so on (if any). This enables the model to generate super-resolved realizations that remain consistent with the underlying cosmological dynamics while accurately reproducing the statistical properties of the simulated fields, whether for cosmological field evolution or super-resolution.

\section{Methods}\label{method}
\subsection{Initial Conditions}\label{IC}
Initial conditions are generated at z=127 (corrresponding to a scale factor of $a \approx 0.0078)$ using \textsc{AxionCAMB} \citep{2015_Hlozek} with an axion mass of $m = 2.5 \times 10^{-22}$eV, assuming all matter is axion dark matter. The resulting power spectra are passed to \textsc{JaxPM}\footnote{\url{https://github.com/DifferentiableUniverseInitiative/JaxPM}}, which generates particle positions and velocities using second-order Lagrangian Perturbation Theory (2LPT). A uniform-grid density field is then constructed via Cloud-in-Cell (CIC) interpolation of the displacement field.
The 2LPT velocity field is used to initialise the phase of the wavefunction. Taking the divergence of Eq.~(\ref{vs}) gives
\begin{equation}
\nabla^2 S = \frac{m}{\hbar}\nabla \cdot \mathbf{v},
\end{equation}
which is a Poisson equation solved spectrally. The initial wavefunction is then fully determined by
\begin{equation}
|\psi(\mathbf{x})| = \sqrt{\frac{\rho}{m}},
\end{equation}
\begin{equation}
\nabla\mathrm{arg}\left(\psi(\mathbf{x})\right) = \nabla S = \frac{m}{\hbar} \mathbf{v (x)}.
\end{equation}
 \subsection{Simulations} \label{sim}
 All our simulations were generated using the cosmological parameters $\Omega_m =0.27$, $\Omega_b =0$, $\Omega_{\Lambda} =0.73$, $H_0 =100$ $\textrm{km}\ \textrm{s}^{-1}\textrm{Mpc}^{-1}$ ($h=1$; set to 1 so that units are in $ h^{-1}$),
\textrm{and} \, $\sigma_8 =0.8$ with ICs as described in section \ref{IC}. Multiple realizations of the initial conditions at z = 127 were generated using different random seeds while fixing the underlying cosmology and initial matter power spectrum. The simulation box size is set to $L = 1 \, h^{-1}\, \textrm{Mpc}$, a scale at which contributions from the quantum pressure term are non-negligible and must be accounted for. All the simulations are generated using the \textsc{Jaxion} \citep{Mocz_Jaxion_2025} code. 

\paragraph{Normalization}

The training of neural networks is significantly improved when the input variables are of order unity. As noted in several previous studies \citep{Thiele_2020,Wadekar_2021, Bernardini_2022}, the choice of normalization scheme plays a crucial role in accurately predicting cosmic fields and reproducing their statistical properties.

With $\psi = \textrm{R} + i\,\textrm{I}$, where $\textrm{R}=\operatorname{Re}(\psi)$ and $\textrm{I}=\operatorname{Im}(\psi)$, define  $D=\{\textrm{R},\textrm{I},V\}$  as the set of fields under consideration. Let $D_{\mathrm{max}}$ and $D_{\mathrm{min}}$ be the global maximum and minimum values computed over the entire dataset (training and test samples combined). We normalize each field through the following sequence of transformations:
\begin{align}
D &\rightarrow \sinh^{-1}(D), \\
D &\rightarrow \frac{D-D_{\mathrm{min}}}
{D_{\mathrm{max}}-D_{\mathrm{min}}}, \\
D &\rightarrow \frac{D-0.5}{0.5},
\end{align}
such that the final normalized values satisfy $D \in [-1,1]$.

Similarly, for the scale factor $a$, we apply
\begin{align}
a &\rightarrow \log_{10}(a), \\
a &\rightarrow \frac{a-a_{\mathrm{min}}}
{a_{\mathrm{max}}-a_{\mathrm{min}}}, \\
a &\rightarrow \frac{a-0.5}{0.5},
\end{align}
so that the normalized scale factor also lies in the interval $[-1,1]$.

\paragraph{Training and Test split}
For evolution task, a single realization of the initial condition was evolved from z = 127 ($a\approx 0.0078$) to z = 5.36 ($a\approx 0.1572$) with a $\Delta a = 10^{-5}$ and the fine snapshots were stored only for the scale factors range $a\in [0.149, 0.157]$, which constitutes total of 796 snapshots out of which 20\% (159 of them) are used for training and rest as the test samples. Thus, the training and test sets correspond to different scale factors from the same realization of the initial condition. Restricting the analysis to a narrow interval in scale factor allows the network to learn the local temporal evolution while providing a dense sampling of closely spaced states. Such a fine resolution is also essential for the simulations to produce a low enough (of the order 1) SP PDE residuals, in accordance with Eqs. (\ref{PE} \& \ref{SSE}). Each snapshot consists of the real and imaginary parts of the wavefunction $\psi$ along with a potential $V$ corresponding to a given scale factor $a$.

For the super-resolution task, we create 80 different realizations of the initial FDM power spectrum. 80\% of these are taken for training and 20 are reserved for testing. For each of those realization, the simulations were evolved from z = 127 to z= 11.45 ($a\approx 0.0803$)  with  $\Delta a = 10^{-6}$, for improving the agreement of the numerical simulations with the PDE residuals, and the fine snapshots were stored only for a scale factors ranging $a_{multi} \in [0.0798, 0.0803]$ constituting 51 snapshots for each realization, giving us a total of 3264 snapshots for both training and test. Unlike the evolution task, the super-resolution model learns a mapping between low- and high-resolution fields at a fixed epoch rather than temporal evolution. Consequently, a large number of statistically independent realizations is more important than sampling a broad range of scale factors. The earlier redshift also reduces the computational cost associated with generating the high-resolution training data while retaining the characteristic FDM structures required for the reconstruction task. We evaluated several downsampling methods, each of which has its own limitations. The key consideration is to select one method and apply it consistently throughout the analysis. In our case, we use zoom\footnote{\href{https://docs.scipy.org/doc/scipy/reference/generated/scipy.ndimage.zoom.html}{SciPy ndimage.zoom documentation}} from scipy with order 1 and mode `nearest'.

\subsection{Cosmo-SPINN}\label{CS_}
\begin{figure*}
    \centering
    \includegraphics[width=\linewidth]{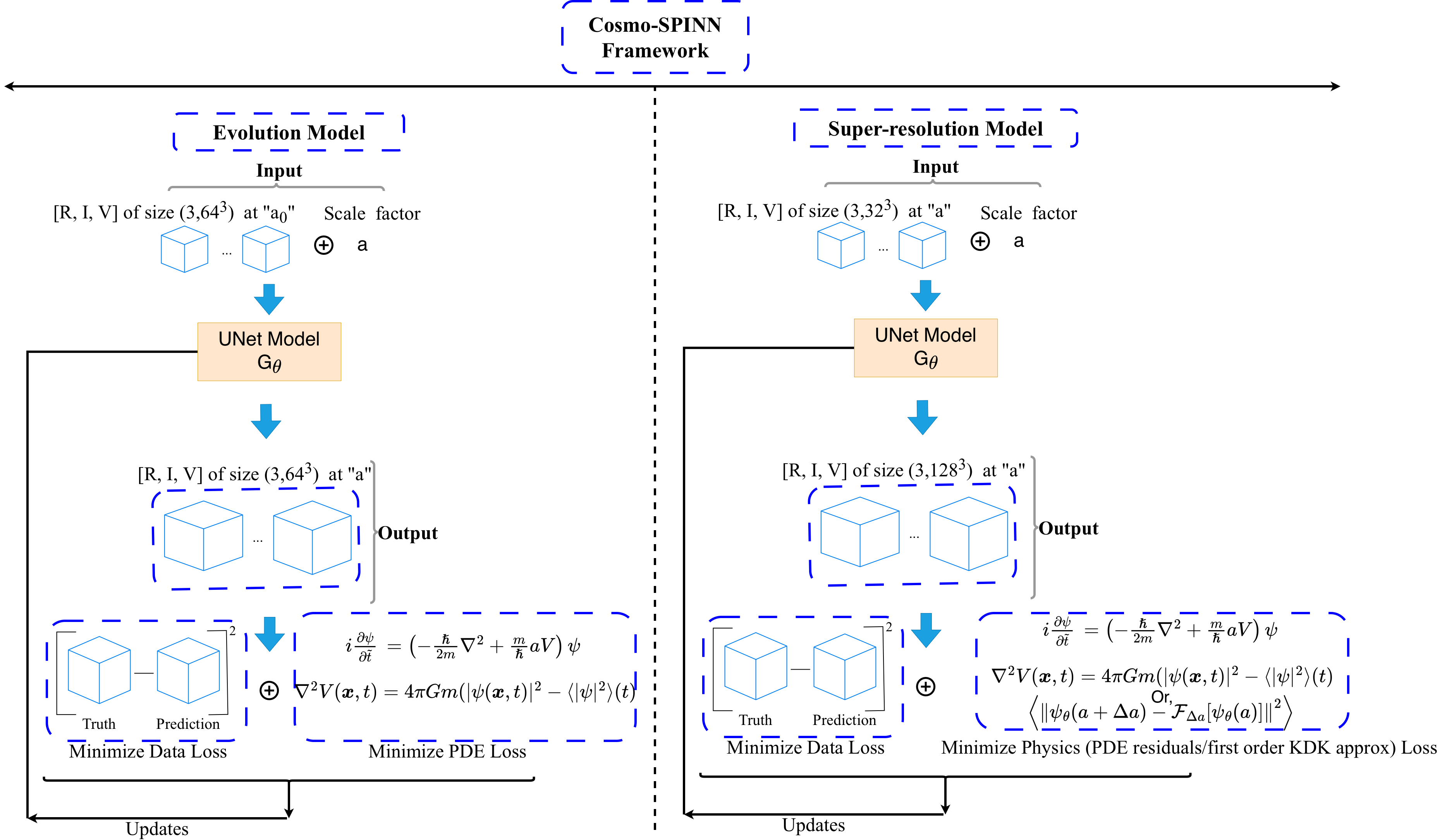}
    \caption{Schematic of the Cosmo-SPINN framework applied to two tasks: (a) \textbf{(left) Evolution Model}, which predicts {R,I,V} ($\text{Re}(\Psi) = \text{R},\text{Im}(\Psi)=\text{I}$) at a target scale factor a from the initial conditions, and (b) \textbf{(right) Super-Resolution Model}, which reconstructs high-resolution fields from low-resolution inputs at the same scale factor. In both cases, the outputs serve as approximate solutions to the SP system and are constrained through a combination of data and physics-informed losses. During training, the parameters $\theta$ are iteratively optimized to obtain $\theta^{*}$, to generate the  required cosmological cubes for wave function and the potential at the specified cosmological scale factor `$a$'}
    \label{fig1}
\end{figure*}

Here we introduce Cosmo-SPINN, a physics-informed generative framework for FDM inside the periodic cubic box of length L = 1$\,h^{-1}$Mpc. As illustrated in Fig.~\ref{fig1}, the framework is applied to two related tasks: (i) the evolution of cosmological fields from initial conditions to a specified scale factor and (ii) the super-resolution of FDM simulations. In both cases, the network predicts the real and imaginary components of the wavefunction together with the gravitational potential while enforcing consistency with the underlying SP dynamics through a physics-informed loss function.

We use $\lambda = \hbar/m$ to simplify notation. Writing out the real and imaginary components of the wavefunction $\psi$ explicitly as $\psi = \text{Re}(\psi) + i \,\text{Im}(\psi) = \textrm{R} + i \, \textrm{I}$, we can rewrite Eqs. (\ref{PE} \& \ref{SSE}) as:
\begin{align}
 \partial_{\tilde{t}} \mathrm{R} &=
 - \frac{\lambda}{2} \nabla^2 \mathrm{I}
 + \frac{1}{\lambda}\tilde{V}\,\mathrm{I}
 \label{eq:sp_re}\\
\partial_{\tilde{t}} \mathrm{I} &=
 \frac{\lambda}{2}\nabla^2 \mathrm{R}
 - \frac{1}{\lambda}\tilde{V}\,\mathrm{R}
 \label{eq:sp_im}\\
 \nabla^2 (\tilde{V}/a) &=
 4\pi G (\rho-\langle\rho\rangle)
 \label{eq:sp_poisson}
\end{align}
As discussed in Section~\ref{GE}, our choice to train the Cosmo-SPINN using the SP equations is motivated by their ability to capture the wave-like behavior inherent to FDM dynamics.
Consequently, we adopt the SP formulation for training our physics-informed model. Specifically, we design the network to predict the real $\mathrm{R}$ and imaginary $\mathrm{I}$ components of the wavefunction $\Psi$, as well as the gravitational potential $V$, which is treated as an auxiliary output rather than computed through direct numerical solution. This strategy can significantly reduce the computational cost with automatic differentiation by circumventing the need to solve Poisson’s equation at every training step, as this is typically solved with FFTs and computing the Fourier transform for an $N \times N$ grid requires $\mathcal{O}(N^2logN)$ operations and approximately $\mathcal{O}(N^2)$ memory, which can quickly be restrictive for larger grids such as in the super-resolution task. However, automatic differentiation requires the whole grid to be passed to the model which can be memory-intensive in case of a single GPU. As a result, we still resort to FFTs for derivatives' calculations in our case.

We define a neural network $G(X;\theta)$ that predicts the real and imaginary components of the wavefunction together with the gravitational potential,
\begin{align}
G(X;\theta)=\left(\textrm{R}_{\theta}(X),\textrm{I}_{\theta}(X),V_{\theta}(X)\right),
\end{align}
where $(\cdot)_{\theta}$ denotes the approximation realized by the network parameters $\theta$.
The precise form of the input $X$ depends on the task considered. For the \textit{ evolution} model, the input consists of the initial conditions together with a target scale factor,
$$
X_{\rm evo}={\psi(a_0),V(a_0),a},
$$
and the network predicts the fields at scale factor $a$. For the\textit{ super-resolution} model, the input consists of low-resolution fields and the corresponding scale factor,
$$
X_{\rm SR}={\psi_{\rm LR}(a),V_{\rm LR}(a),a},
$$
and the network predicts the corresponding high-resolution fields. A schematic overview of both tasks is shown in Fig.~\ref{fig1}.
We predict real $\mathrm{R}$ and imaginary $\mathrm{I}$ parts of $\psi$ directly along with the potential $V$. This approach ensures mass conservation and correct evolution, eliminating the need to explicitly implement a continuity equation.% A schematic of the same is shown in Fig. (\ref{fig1}).

The data loss is defined as the mean squared error (MSE) between the predicted and target fields,
\begin{equation}
\mathcal{L}_{\mathrm{data}}
=
\frac{1}{N_{\mathrm{elem}}}
\sum_{j=1}^{N_{\mathrm{elem}}}
\left(
y_{{\rm pred},j}
-
y_{{\rm true},j}
\right)^2,
\end{equation}
where $y=(\mathrm{R},\mathrm{I},V)$ contains the real and imaginary components of the wavefunction together with the gravitational potential, and $N_{\mathrm{elem}}$ is the total number of tensor elements (all spatial grid points and all field channels).
The total loss is defined as
\begin{equation}
    \mathcal{L}_{\mathrm{total}} = \mathcal{L}_{\mathrm{data}} + \lambda_w\,\mathcal{L}_{\mathrm{physics}},
\end{equation}
where $\lambda_w$ is a weighting coefficient, which is chosen such that both the losses contribute equally in the gradient space. We explore two formulations of the physics loss $\mathcal{L}_{\mathrm{physics}}$.

\paragraph{Method 1: PDE Residuals.}
A batch of scale factors $a$ are drawn uniformly at random from the range spanned by the training data, namely $a \in[0.149,0.157]$ for the evolution model and $a\in[0.0798,0.0803]$ for the super-resolution model. The generator is queried at each sampled $a$ and at the corresponding $a + \Delta a$, and the PDE residuals are evaluated from the predicted fields. For a $(d+1)$-dimensional system, the residuals associated with Eqs.~\eqref{eq:sp_re}--\eqref{eq:sp_poisson} are
\begin{align}
\label{r1}
\mathcal{R}_{\mathrm{R}}(X) &= \partial_{\tilde{t}}\mathrm{R}_{\theta} + \frac{\lambda}{2}\!\left(\sum_{i=1}^{d}\partial_{x_i}^2\mathrm{I}_{\theta}\right) - \frac{a}{\lambda}V_{\theta}\,\mathrm{I}_{\theta}, \\
\label{r2}
\mathcal{R}_{\mathrm{I}}(X) &= \partial_{\tilde{t}}\mathrm{I}_{\theta} - \frac{\lambda}{2}\!\left(\sum_{i=1}^{d}\partial_{x_i}^2\mathrm{R}_{\theta}\right) + \frac{a}{\lambda}V_{\theta}\,\mathrm{R}_{\theta}, \\
\label{r3}
\mathcal{R}_V(X) &= \sum_{i=1}^{d}\partial_{x_i}^2 V_{\theta} - 4\pi G\left(\rho -\langle \rho \rangle\right).
\end{align}
The Laplacians here are computed with the help of FFT methods and the time derivative is evaluated using first order finite difference method. The physics loss is then taken as the mean squared residual,
\begin{equation}
    \mathcal{L}_{\mathrm{physics}}^{(1)} = \left\langle\mathcal{R}_{\mathrm{R}}^2 + \mathcal{R}_{\mathrm{I}}^2 + \mathcal{R}_V^2\right\rangle.
\end{equation}

\paragraph{Method 2: Kick--Drift--Kick Approximation.}
As in Method 1, a batch of scale factors $a$ are drawn uniformly at random from the corresponding training interval. 
The generator predicts a field at $a$, which is then evolved forward by 
one step $\Delta a$ using a first-order approximation of the kick--drift--kick 
integrator. The generator is also queried directly at $a + \Delta a$, and the 
physics loss is the residual between the two,
\begin{equation}
    \mathcal{L}_{\mathrm{physics}}^{(2)} = \left\langle\left\|\psi_{\theta}(a+\Delta a) 
    - \mathcal{F}_{\Delta a}\!\left[\psi_{\theta}(a)\right]\right\|^2\right\rangle,
\end{equation}
where $\mathcal{F}_{\Delta a}$ denotes one step of the numerical integrator. 
Expanding each exponential operator in the KDK scheme to first order, 
$e^{\hat{A}} \approx 1 + \hat{A}$, the evolved wavefunction is approximated as
\begin{align}
\mathcal{F}_{\Delta a}\!\left[\psi\right] \approx\;
&\left(1 - i\frac{m}{\hbar}\frac{\Delta t}{2a}V(t+\Delta t,\mathbf{x})\right)
\nonumber\\[4pt]
&\times\left(1 + i\frac{\hbar}{2m}\frac{\Delta t}{a^2}\nabla^2\right)
\nonumber\\[4pt]
&\times\left(1 - i\frac{m}{\hbar}\frac{\Delta t}{2a}V(t,\mathbf{x})\right)\psi(t,\mathbf{x}).
\end{align}
This linearisation yields a computationally tractable residual that enforces consistency between the generator output and the numerical integrator without requiring full operator exponentiation at training time.
The solution is then realized by minimizing the total loss ($\mathcal{L}_{\textrm{data}}+\lambda_w \, \mathcal{L}_{\textrm{physics}}$) through the optimization of the neural network defined by $\theta$:
\begin{align}
    \theta^* = \underset{\theta}{\text{arg min}} \left(\mathcal{L}_{\textrm{data}}+\lambda_w \, \mathcal{L}_{\textrm{physics}}\right)
\end{align}
We do not include any explicit constraints such as mass conservation, or periodic boundary conditions as these are obeyed when the network is sufficiently optimized.

The network training procedure  largely follows that of conventional neural networks, employing automatic differentiation~\citep{Auto_diff} to compute gradients and backpropagation~\citep{DL_Lecun} for parameter optimization. The key difference is the incorporation of an additional physics-informed loss term that enforces consistency with the governing equations.
\begin{figure*}
\centering
\hspace*{-1.2cm}
\begin{tabular}{cc}
\includegraphics[width=0.5\linewidth]{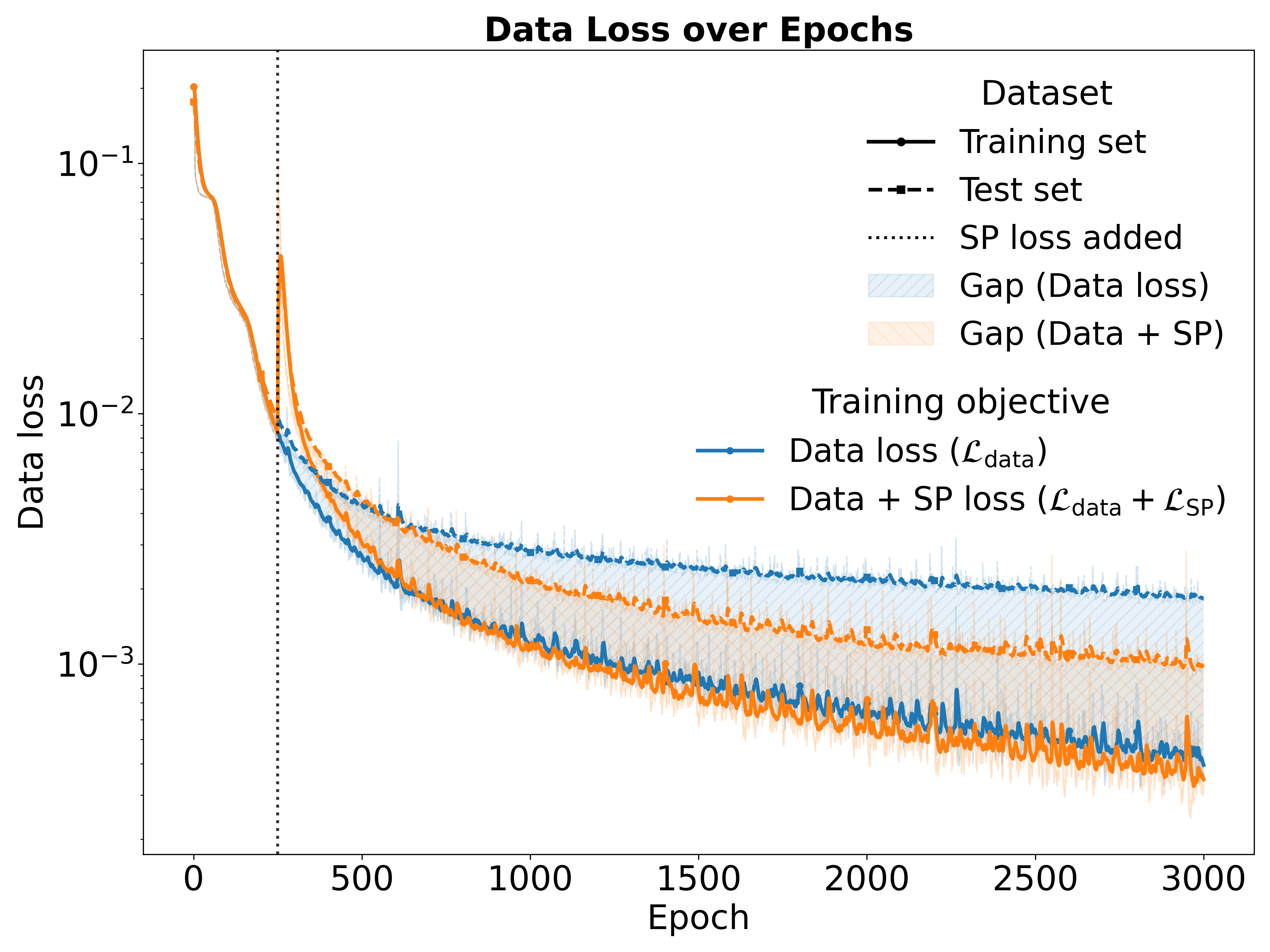}&
\includegraphics[width=0.5\linewidth]{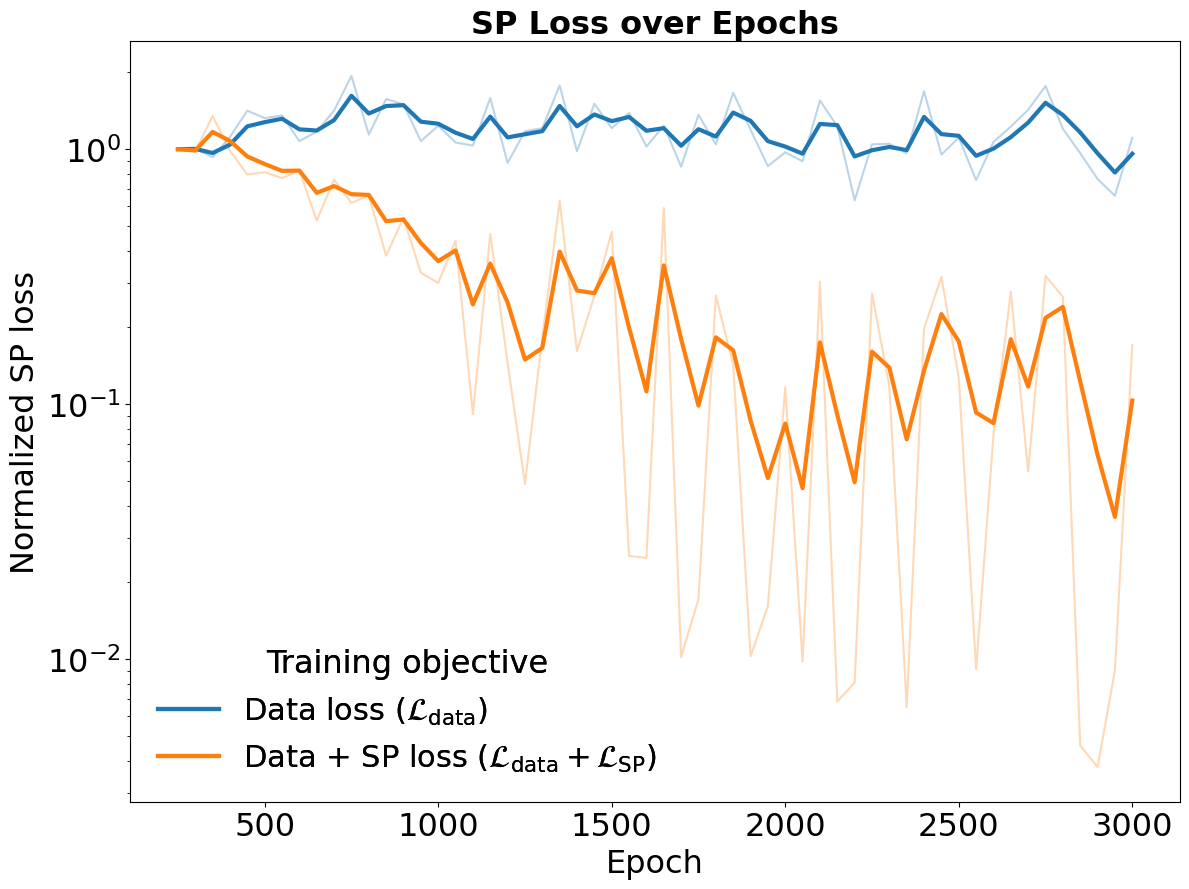}
\end{tabular}
\caption{Training and test data losses (left) and normalized SP loss (right) for the evolution model. Training is performed using only the data loss for the first 250 epochs, after which the SP loss is added to the objective (vertical dotted line) and optimization continues for the remaining 2750 epochs. Solid and dashed curves denote the training and test losses (shown losses are smoothed with an exponential moving average to remove small-scale variability in the faint raw curves shown in the background), respectively, while the shaded regions indicate the train--test gap. Blue curves correspond to the data-only baseline, whereas orange curves correspond to the physics-informed model initialized from the same 250-epoch checkpoint. Incorporating the SP loss improves both the SP objective and the data reconstruction performance.}
\label{fig2}
\end{figure*}

\subsection{Architecture and Optimization}

Both models share a common set of building blocks. Scalar conditions 
(scale factor $a$) are encoded via a Fourier feature embedding,
\begin{equation}
    \gamma(a) = \left[\sin\!\left(2^k \pi\, a\right),\, 
    \cos\!\left(2^k \pi\, a\right)\right]_{k=0}^{K-1} \in \mathbb{R}^{2K},
\end{equation}
with $K=8$, producing a 16-dimensional embedding that is passed through 
a two-layer MLP to yield a $c$-dimensional conditioning vector (c = 64 for us). This 
vector is spatially broadcast and concatenated channel-wise to the input 
volume before the first convolution. Spatial feature extraction is 
performed by 3D ResNet blocks, each consisting of two $3\times3\times3$ 
convolutions with LeakyReLU activations and a skip connection; when the 
number of channels changes, the skip connection uses a $1\times1\times1$ 
convolution, otherwise it is the identity.

\paragraph{Evolution.}
The evolution generator follows a U-Net-style encoder--decoder with 
skip connections. The input is a $3\times64^3$ field concatenated with 
the spatially expanded conditioning vector, giving a $(3+c)\times64^3$ 
tensor. An initial $3\times3\times3$ convolution projects this to 64 
channels. The encoder applies one ResNet block at $64^3$ resolution, 
followed by a strided convolution ($\mathrm{stride}=2$) to $32^3$ with 
128 channels, another ResNet block, a second strided convolution to 
$16^3$ with 256 channels, and a final ResNet block. A two-block 
bottleneck at $16^3$ further processes the feature map. The decoder 
mirrors the encoder: two transposed convolutions ($\mathrm{stride}=2$) 
progressively restore the resolution to $32^3$ and $64^3$, with skip 
connections from the corresponding encoder stages concatenated 
channel-wise before each decoder ResNet block. A final ResNet block and 
$1\times1\times1$ convolution produce a $3\times64^3$ correction 
$\delta$. The output is formed as a residual,
\begin{equation}\label{Ev_1}
    \hat{\psi}(a) = \psi(a_0) + 0.1\,\delta,
\end{equation}
so the network learns a small correction to the input field rather than 
predicting the full output from scratch.

\paragraph{Super-Resolution.}
The super-resolution generator upsamples a low-resolution $3\times32^3$ 
input to $3\times128^3$. The input is concatenated with the conditioning 
vector and projected to 64 channels by an initial convolution. Two 
ResNet blocks process the features at the input $32^3$ resolution, after 
which two successive transposed convolutions ($\mathrm{stride}=2$) 
upsample to $64^3$ and $128^3$, each followed by two ResNet blocks at 
the new resolution. A $1\times1\times1$ convolution produces the 
high-resolution residual, which is added to a trilinearly interpolated 
($\times4$) version of the low-resolution input,
\begin{equation}
    \hat{\psi}_{\mathrm{HR}} = \mathrm{Interp}_{\times 4}\!\left[
    \psi_{\mathrm{LR}}\right] + \delta_{\mathrm{HR}},
\end{equation}
ensuring the coarse-scale structure is preserved while the network 
focuses on recovering fine-scale features.

 Both networks are optimised using Adam \citep{2014_Kingma} with a learning rate of 
$10^{-4}$ and momentum parameters $\beta_1 = 0.9$, $\beta_2 = 0.999$. 
Evolution model is trained with a batch size of 16, while for the 
super-resolution network the $128^3$ output volume imposes significant 
memory demands, restricting training to a batch size of 2 on a single 
GPU. Both models are trained in two stages: an initial warm-up phase on 
the data loss $\mathcal{L}_{\mathrm{data}}$ alone, followed by a second 
phase in which the physics loss $\mathcal{L}_{\mathrm{physics}}$ is 
added. For evolution network, the two stages span 250 and 2750 epochs 
respectively; for the super-resolution model, 20 and 60 epochs 
respectively.
\section{Results}\label{Res}
\subsection{Training Convergence and Physics Consistency of Models}
\begin{figure*}
\centering
\hspace*{-1.2cm}
\begin{tabular}{cc}
\includegraphics[width=0.5\linewidth]{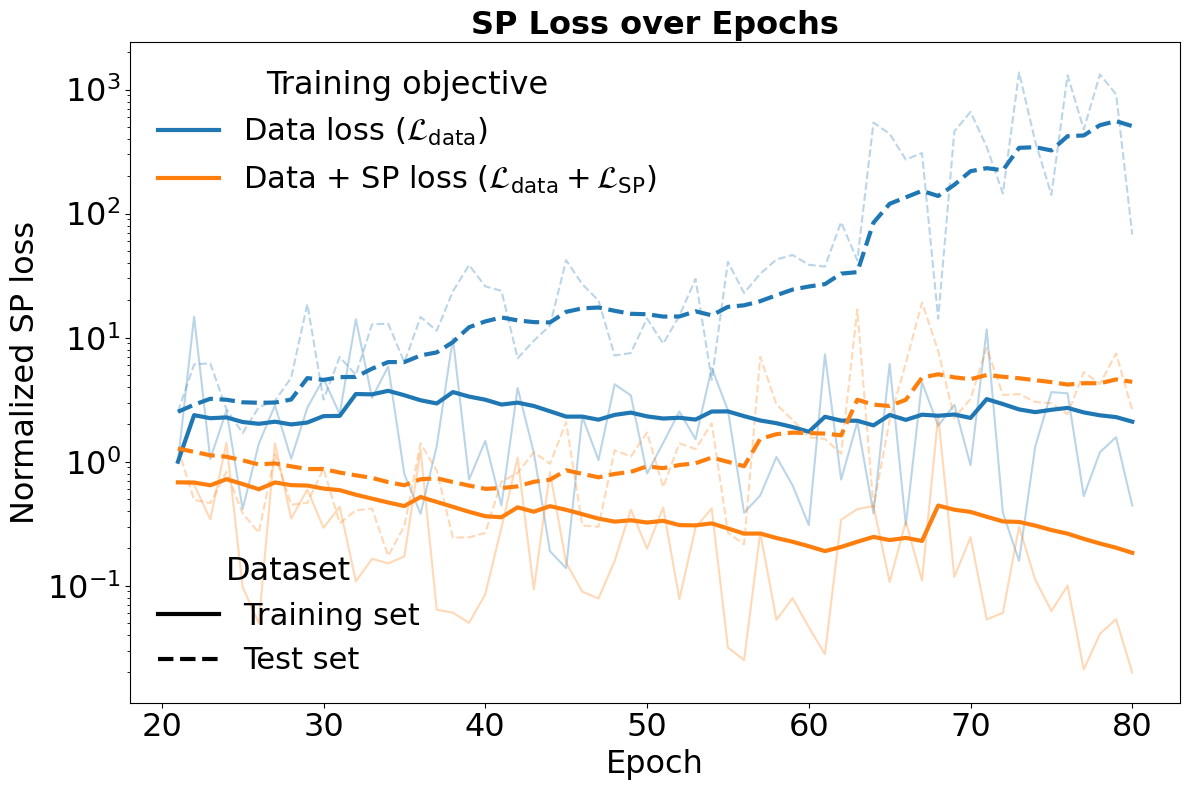}&
\includegraphics[width=0.5\linewidth]{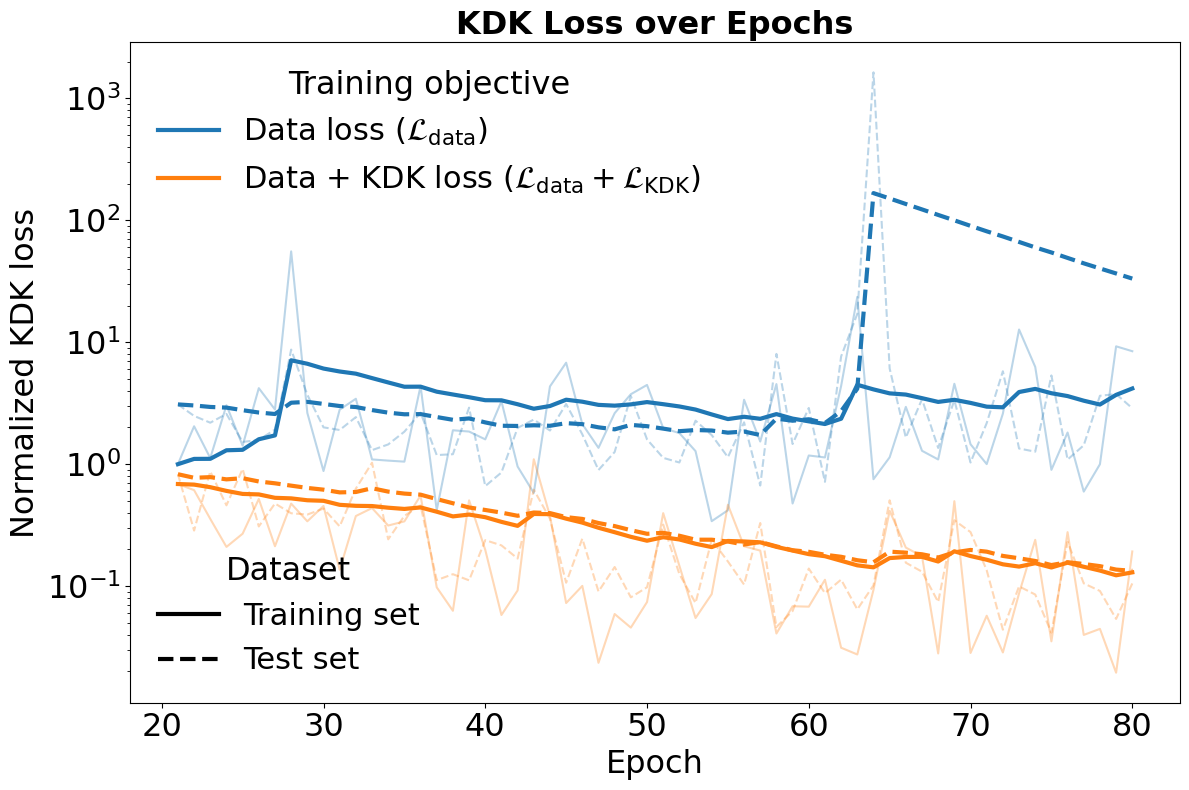}
\end{tabular}
\caption{Evolution of the normalized physics loss for the super-resolution model. Left: physics loss based on the SP residual. Right: physics loss based on a first-order KDK approximation. Physics-informed training significantly reduces the losses on both the training and test sets. Solid and dashed lines denote the training and test sets, respectively (shown losses are
smoothed with an exponential moving average with $\alpha$= 0.1 to remove small-scale variability in the faint raw curves shown in the background). Blue curves correspond to models trained using only the data (MSE) loss, while orange curves correspond to models trained using the data loss together with the SP (left) or KDK (right) physics loss, introduced after epoch 20 (the first 20 epochs use only MSE data loss). }
\label{fig3}
\end{figure*}

We first evaluate the convergence behaviour of the Evolution Model shown in Fig.~\ref{fig1} (left). Figure~\ref{fig2} shows the evolution of the training and test data losses (left panel) alongside the normalized SP loss (right panel). We consider two different traning strategies: (i) using only data loss, and (ii) using data loss and SP loss after epoch 250. After the physics-informed term is introduced at epoch 250, the optimization dynamics change as the network is required to satisfy both data fidelity and physical consistency. While the normalized SP loss decreases by nearly an order of magnitude for the physics-informed model, it remains nearly constant for the data-only baseline throughout the remainder of training, indicating that supervised optimization alone is insufficient to improve consistency with the underlying physical dynamics described by the SP equations. At the same time, both the training and test data loss continue to decrease while remaining comparable to those obtained using the data-only objective. Furthermore, the reduced train--test gap suggests that the SP constraint acts as a useful regularizer, guiding the model toward physically admissible solutions that also generalize better to unseen data.

The combined objective here is defined as
\begin{align}
    \mathcal{L}
    =
    \mathcal{L}_{\mathrm{data}}
    +
    \alpha(t)\,\lambda_w\,\mathcal{L}_{\mathrm{SP}},
\end{align}
where $\lambda_w$ is an adaptive gradient-balancing factor that equalizes the gradient magnitudes of the data and SP losses, while $\alpha(t)$ controls the overall strength of the physics constraint. Rather than introducing the physics loss abruptly, its contribution is gradually increased according to
\begin{align}
    \alpha(t)=\alpha_{\max}\min\left(1,\frac{t}{N_{\mathrm{ramp}}}\right),
\end{align}
where $\alpha_{\max}=0.02$ and $N_{\mathrm{ramp}}=1000$ epochs. This warm-up strategy allows the network to first learn an accurate data-driven representation before progressively enforcing the physical constraint, thereby avoiding the optimization instabilities associated with suddenly introducing a strong additional objective.

Although the adaptive weighting balances the optimization in gradient space, the SP loss remains a considerably stronger constraint since it is evaluated in physical space and directly enforces the underlying nonlinear dynamics. Moreover, the evolution model is trained using a single realization of the initial conditions, for which 796 simulation snapshots are available over the interval $a\in[0.0798,\,0.0803]$. Only $20\%$ of these redshifts are used for supervised training, whereas the SP loss is evaluated on randomly sampled redshifts spanning the entire interval. Consequently, the physics objective constrains a substantially larger portion of the evolution than the supervised data loss. Without the additional scaling and gradual ramp-up, the optimization becomes biased toward satisfying the governing equations at the expense of accurately reconstructing the available training snapshots. We therefore set $\alpha_{\max}=0.02$, which provides a suitable balance between physical consistency and data fidelity. Empirically, larger values of $\alpha_{\max}$ further reduce the SP residual and improve the test loss, but lead to a noticeable deterioration in the reconstruction of the training data.

Unlike first-order time-stepping constraints such as KDK, the SP loss directly constrains the underlying dynamics rather than enforcing agreement over a single numerical integration step. Our objective is therefore to incorporate the full physical dynamics into the learning process whenever possible. Nevertheless, the increased complexity of the SP constraint makes the optimization more challenging, motivating us to investigate alternative formulations of the physics loss for more demanding learning tasks.

\begin{figure*}
    \centering
    \includegraphics[width=\linewidth]{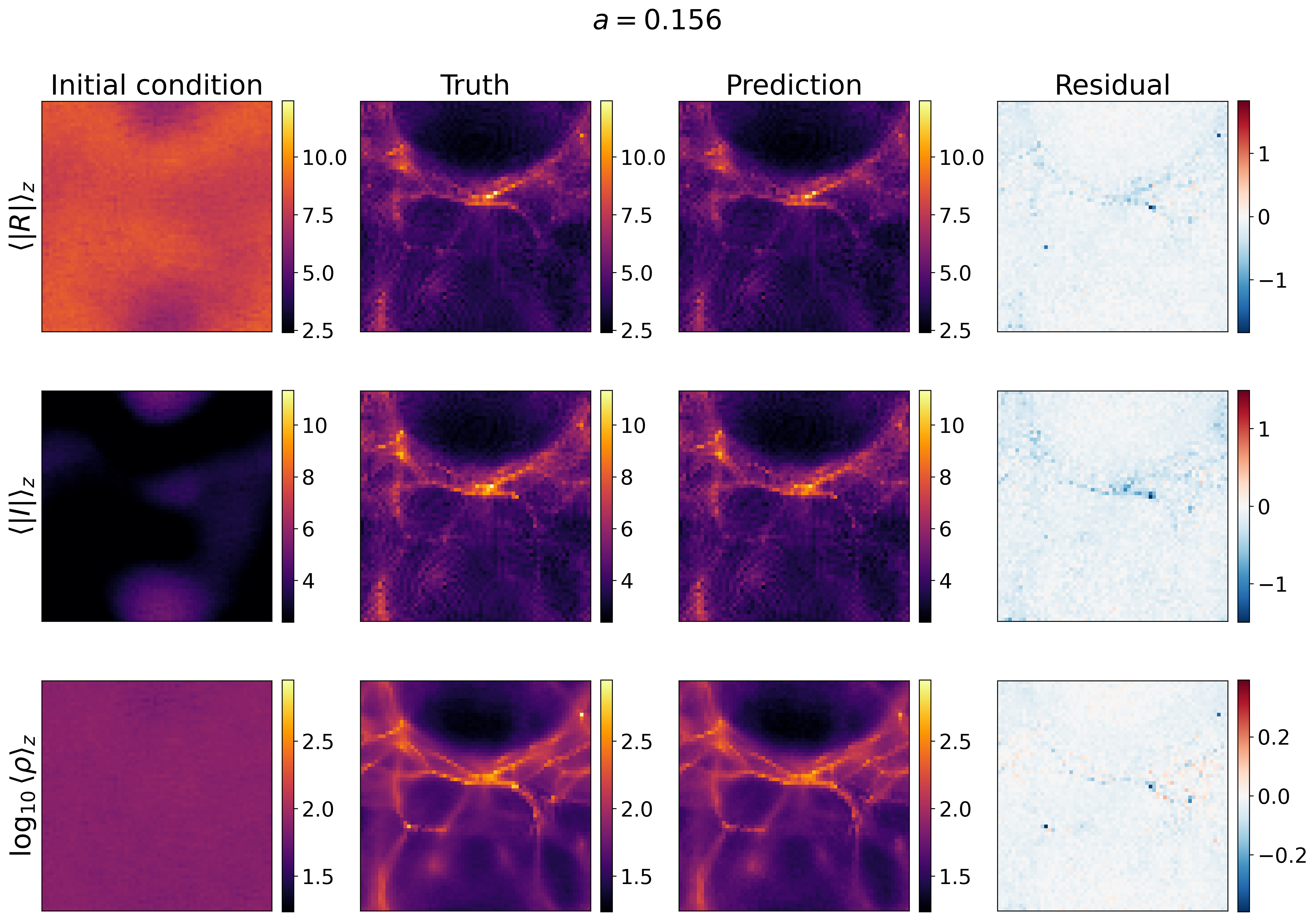}
    \caption{Evolution Model: Comparison of the ground-truth and predicted projected fields $\langle |R|\rangle_z$, $\langle |I|\rangle_z$, and density log$_{10}\langle \rho\rangle_z$ at scale factor a=0.156, obtained by averaging the corresponding $64^3$ fields along the $z$-axis. Columns show the initial condition, reference solution, neural-network prediction, and residual (prediction-truth). The predicted wavefunction components and density closely match the reference solution, with residuals primarily confined to dense filamentary structures and other small-scale features.}
    \label{fig4}
\end{figure*}

We therefore compare two different formulations of the physics loss for the super-resolution model shown in Fig. \ref{fig1} (right): the SP residual and a first-order KDK approximation, as described in Section \ref{CS_}. Unlike the evolution model, no additional scaling parameter $\alpha$ is introduced in this case. The super-resolution model is trained using all available redshifts, with the train--test split performed over independent realizations rather than temporal snapshots. Consequently, both the supervised and physics losses are evaluated over the same range of redshifts, eliminating the imbalance present in the evolution model. Our objective here is therefore to investigate whether an appropriate formulation of the physics loss can provide stable optimization without requiring an additional stabilization parameter such as $\alpha$.

For this, the super-resolution model is first trained for the initial 20 epochs using only the MSE data loss before introducing the respective physics losses. A baseline model trained with only MSE objective (data loss) is also included to isolate the effect of the physical regularization, as illustrated in Fig.~\ref{fig3} with the blue curves. Figure~\ref{fig3} shows the evolution of the normalized SP (left panel) and KDK (right panel) losses for both the baseline model (blue curves) and the physics-informed model (orange curves), with training (solid) and test (dashed) curves shown for each. In both cases, the inclusion of the physics loss leads to a significant reduction in the normalized physics losses on both the training and test sets, demonstrating that the model learns solutions that better satisfy the underlying dynamics. However, the two formulations of the physics loss exhibit noticeably different convergence behaviour. For the SP-based loss, the training residual decreases steadily, but after approximately 40 epochs the test residual begins to diverge from the training residual, regardless of whether the physics loss is included. The primary effect of the physics loss in this case is to reduce the overall magnitude of the residual rather than eliminate the growing train--test gap. This behaviour likely reflects the increased complexity of the coupled SP equations. In contrast, the KDK-based constraint produces a much smoother and more stable convergence. The training and test residuals closely track one another throughout training, indicating improved stability and better generalization to unseen realizations and redshift slices. Although the final training loss is slightly higher than that obtained with the SP formulation, the smaller train--test gap suggests that the KDK approximation provides a more robust physics regularization for this super-resolution task.

Overall, both approaches improve the physical consistency of the generated fields relative to the data-only model. The SP residual achieves a larger reduction in the physics error, while the KDK formulation yields more stable training and better agreement between training and test performance.

\subsection{Assessment of the test samples (Visual inspection and power spectrum)}

We now assess the performance of both models on unseen test data. We first examine the evolution model, evaluating its ability to reproduce the wavefunction and density fields. Additional results demonstrating its generalization to previously unseen realizations of the initial conditions are presented in Appendix~\ref{appendix_B}. We then consider the super-resolution model, which constitutes the primary contribution of this work, comparing reconstructions obtained using the SP-residual and KDK-based physics losses (throughout the remainder of this section, we denote the super-resolution model trained with the SP-residual loss as the SP model and the model trained with the KDK-based loss as the KDK model). All super-resolution results presented in this section are obtained using the model checkpoints selected according to the minimum physics test loss during training shown in Fig.~\ref{fig3}.  Note that, while all models used in this work were trained on the training set discussed in Section \ref{sim}, the snapshots presented here were taken from the reserved test set. That is, the simulation data used to produce the plots that follow were not used to train any of the models.

\subsubsection{Evolution Model}

\paragraph{Evolution Test Samples} Figure~\ref{fig4} demonstrates that the evolution model accurately reproduces the projected wavefunction components, $\langle |R| \rangle_z$ and $\langle |I| \rangle_z$, as well as the projected density field, $\log_{10}(\langle \rho \rangle_z)$, at $a=0.156$. The predicted fields closely match the reference solution, successfully capturing the filamentary network and high-density structures. The residual maps show that the vast majority of pixels remain within a $\sim 5\%$ error level, while the largest deviations are confined to a small number of isolated pixels associated with dense, rapidly varying structures. Even in these regions, the residuals remain below $\sim 10\%$. Such localized discrepancies are expected due to the nonlinear relation $\rho \propto \textrm{R}^2 + \textrm{I}^2$ and the increased sensitivity of high-density regions to small prediction errors. Overall, the model preserves both the large-scale structure and the small-scale features of the underlying SP solution with high accuracy. Despite being trained on only $\sim20\%$ of the available temporal snapshots, the model accurately reproduces the power spectrum across the full training redshift range. The mean relative power-spectrum error remains typically at the $15$--$20\%$ level and below $\sim30\%$ for most scale factors (Appendix~\ref{appendix_A}), indicating stable temporal interpolation and consistent recovery of the underlying dynamics. Further, when trained on multiple realizations, it accurately recovers large scale morphology for unseen test realizations as shown in Appendix~\ref{appendix_B}, Fig. (\ref{fig9}) but suffers in reproducing the small-scale features  for all high k modes. 

%###########################################################################################################

%\paragraph{SR test samples} 

\begin{figure*}
\centering
\hspace*{-0.3cm}
\includegraphics[width = \linewidth]{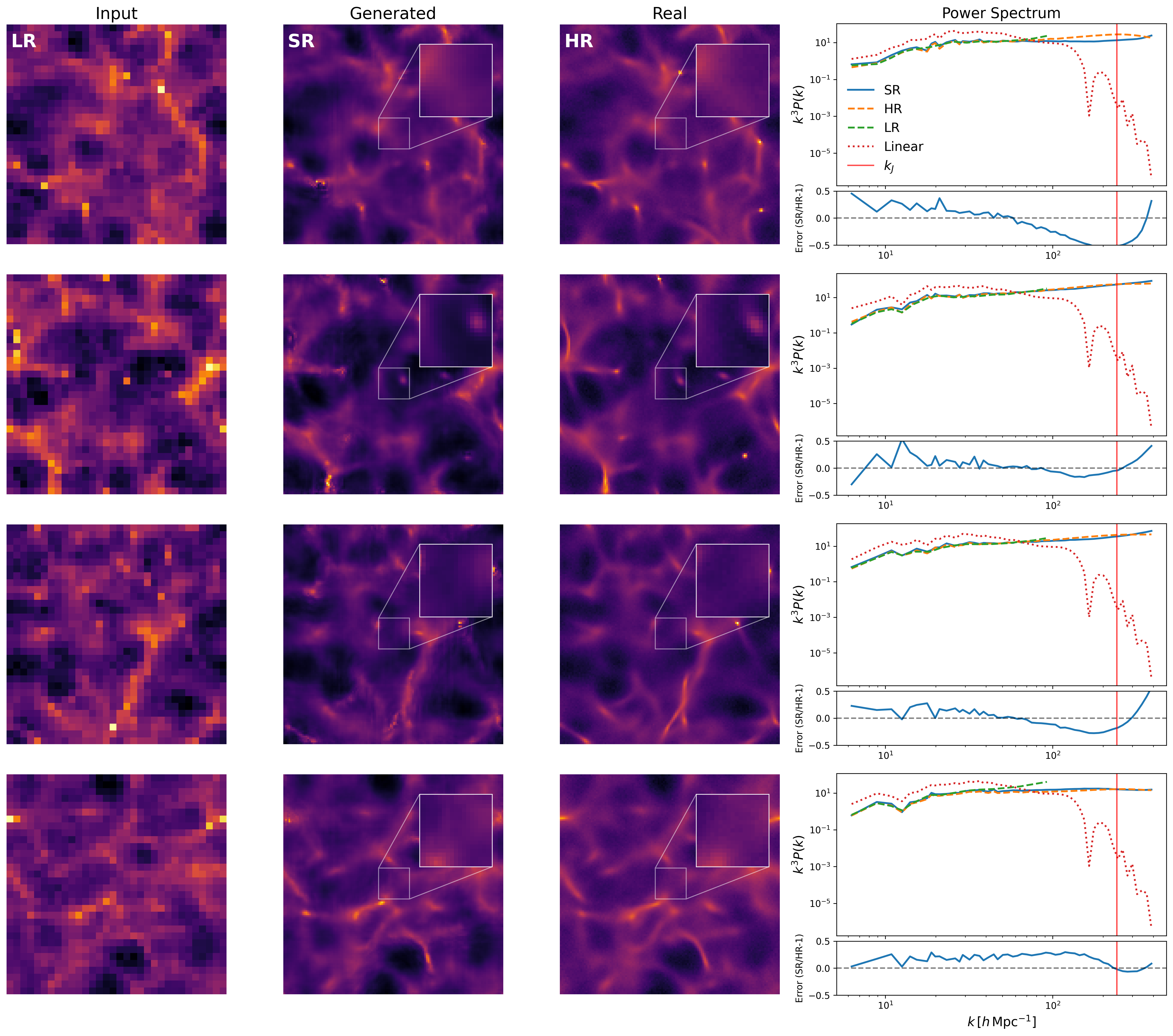}
\caption{Examples of super-resolution density fields  generated by the model trained with the SP physics loss for \textit{the unseen test realizations}. The corresponding scale factors are (from top to bottom) a $\approx$ 0.08039, 0.08042, 0.08043, and 0.08039. The input low-resolution fields ($32^3$, LR), generated super-resolved fields ($128^3$, SR), and target high-resolution fields ($128^3$, HR) are shown as density projections along the z-axis on a logarithmic scale. Insets show zoomed-in regions. The right panels compare the dimensionless power spectra $k^3P(k)$ of the LR, SR, and HR fields, while the lower panels show the fractional error ($P_{SR}/P_{HR}-1$). The vertical red line marks the Jeans-scale wavenumber. The model recovers small-scale structure beyond the input resolution and roughly reproduces the target power spectrum over a wide range of scales, except with some discrepancy at small scales}
\label{fig5} 
\end{figure*}
\begin{figure*}
\centering
\hspace*{-0.3cm}
\includegraphics[width = \linewidth]{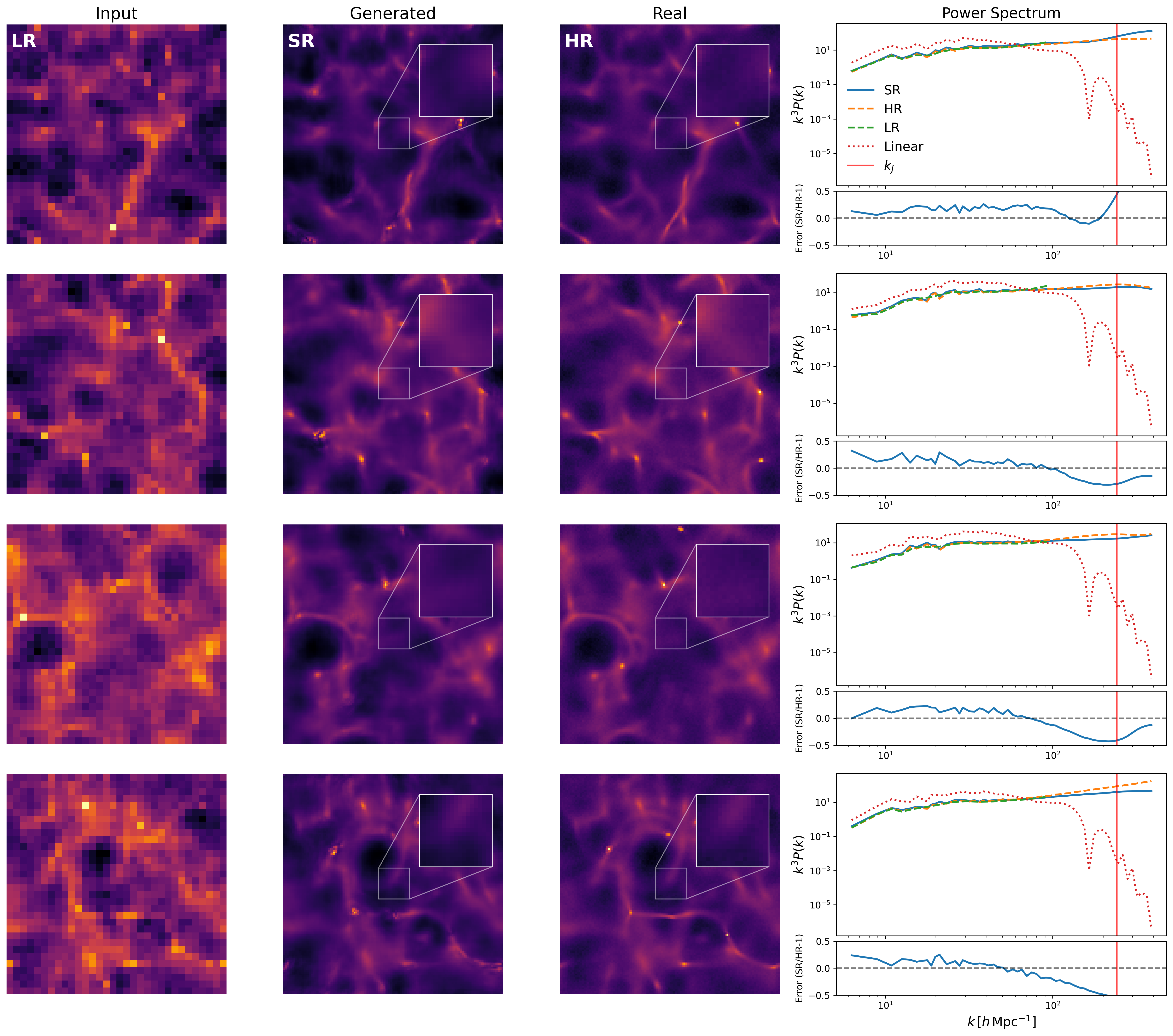}
\caption{Examples of super-resolution density fields  generated by the model trained with the first-order KDK approx. for \textit{the unseen test realizations}. The corresponding scale factors are (from top to bottom)  $a \approx$ 0.08043, 0.08039, 0.08042, and  0.08039. The input low-resolution fields ($32^3$, LR), generated super-resolved fields ($128^3$, SR), and target high-resolution fields ($128^3$, HR) are shown as density projections along the z-axis on a logarithmic scale. Insets show zoomed-in regions. The right panels compare the dimensionless power spectra $k^3P(k)$ of the LR, SR, and HR fields, while the lower panels show the fractional error ($P_{SR}/P_{HR}-1$). The vertical red line marks the Jeans-scale wavenumber. The model recovers small-scale structure beyond the input resolution and roughly reproduces the target power spectrum over a wide range of scales, except with some discrepancy at small scales}
\label{fig6}
\end{figure*}

\subsubsection{Super-Resolution Model}
\paragraph{SR Test Samples}Figures~\ref{fig5} and~\ref{fig6} compare super-resolution (SR) reconstructions obtained using the SP physics loss and the first-order KDK approximation, respectively. The corresponding dimensionless power spectra, ($\Delta(k)=k^3P(k))$, are also shown for the considered test realizations at the selected redshift. The spectra are plotted up to the Nyquist frequency of the grid ($\simeq 402 \, \textrm{h}\, \mathrm{Mpc}^{-1}$), including the LR fields. We also indicate the Jeans wavenumber computed using Eq. \ref{jeans_wn}, ($k_J \simeq 240 \, \textrm{h}\, \mathrm{Mpc}^{-1}$), which marks the scale where quantum pressure becomes important. For ($k < k_J$), gravity dominates the evolution. The fundamental mode is $(k_f \simeq 6 \, \textrm{h}\, \mathrm{Mpc}^{-1})$, implying that approximately 60\% of the resolved modes ($k_f < k < k_J$) lie in the gravity-dominated regime. Also, the linear power spectrum is constructed using the approximate transfer function T(K) given by \citet{Wayne_2000}:
$$T(k) = \frac{\cos{x_J^3}}{1+x_J^8}$$ where, 
\begin{align*}
x_J=1.61m_{22}\, \frac{k}{k_{J,eq}}; k_{J,eq} = 9m_{22}^{1/2} \, \textrm{Mpc}^{-1}\, ;\,\\
m_{22} = m/10^{-22} eV
\end{align*}
 where $m=2.5$ in our implementation. Along with T(k), we also use the initial power spectrum $P_{ic}$ and obtain the linear power spectrum at corresponding scale factor `$a$' as $P_{lin} = (a/a_0)^2 \, T(k)^2 \, P_{ic}$. Both models recover small-scale structures absent from the low-resolution input and reproduce the large-scale morphology of the target high-resolution fields. However, several systematic differences emerge in both the reconstructed density fields and their statistical properties.

Visually, both models recover the main filamentary network, with good agreement in the locations of filaments, nodes, and voids. The SP model produces sharper filaments and higher-contrast density peaks, particularly around compact overdense regions. The zoomed-in regions also reveal a richer small-scale texture. Meanwhile, the KDK model yields smoother reconstructions, with broader filaments and less pronounced density fluctuations. This difference likely reflects the stronger physical constraints imposed by the full SP dynamics, which retain information associated with small-scale density gradients.

The power spectra provide a quantitative assessment of these visual differences. For both models, the SR spectra closely match the HR spectra over a broad range of wavenumbers, demonstrating successful recovery of power beyond the LR resolution limit. However, for the representative realizations shown here, the SP model provides a closer match at high-$k$, remaining consistent with the HR spectrum up to scales approaching $k_J$. The KDK model exhibits a modest suppression of power at high-$k$, consistent with the smoother appearance of the reconstructed fields. While the network accurately reconstructs the large-scale wavefunction, small errors in the real and imaginary components are amplified by the derivative-dependent quantum pressure term. Consequently, discrepancies become most apparent in the quantum-pressure-dominated regime, $(k \gtrsim k_J)$, where the reconstructed spectra begin to deviate from the HR target.

\begin{figure*}
\centering
\hspace*{-0.8cm}
\begin{tabular}{cc}
\includegraphics[width=0.5\linewidth]{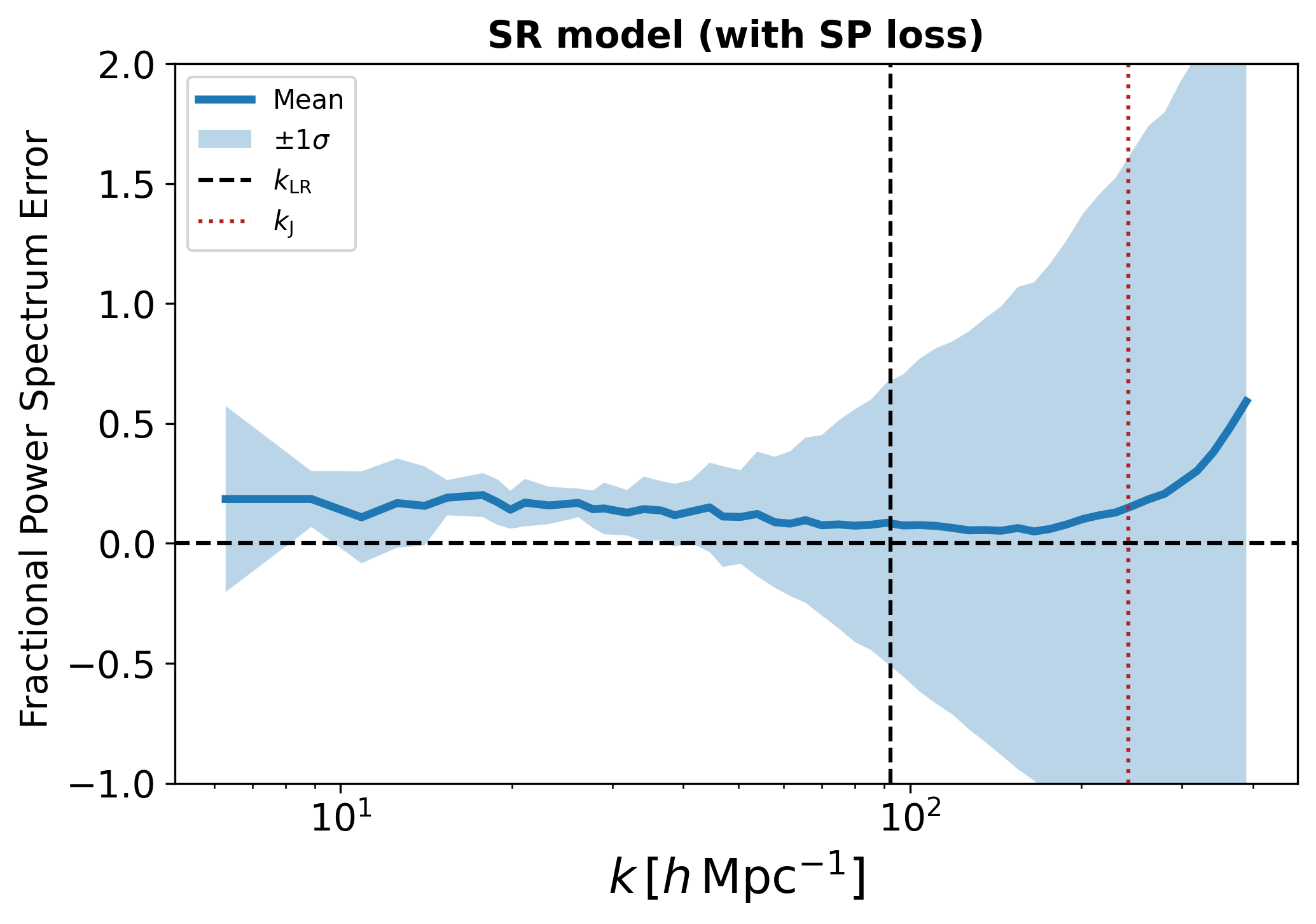}&
\includegraphics[width=0.5\linewidth]{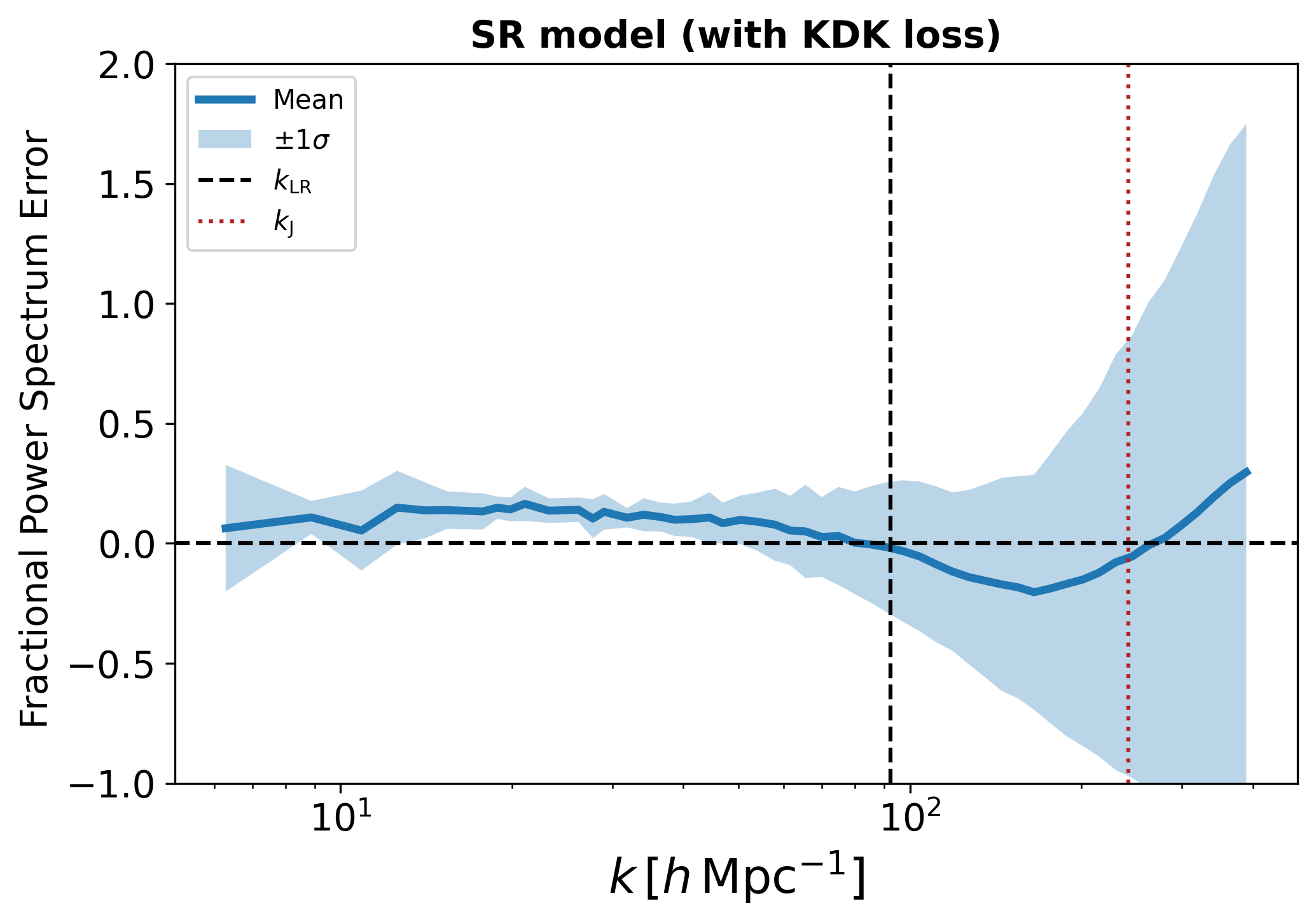}
\end{tabular}
\caption{Mean fractional power-spectrum error, $(P_{\rm SR}/P_{\rm HR}-1)$, across the test realizations at the mid-scale factor $a \approx 0.08041$. The shaded regions denote the corresponding $1\sigma$ scatter across realizations. Left: SR model trained with the SP-residual based physics loss. Right: SR model trained with the first-order KDK-based physics loss.}
\label{fig7}
\end{figure*}

These observations can be interpreted in terms of the physics losses shown in Fig. \ref{fig3}. Although the SP model exhibits a larger test loss than the KDK model, it more accurately reproduces the high-$k$ power spectrum in the representative examples shown here. The two physics objectives emphasize different aspects of the reconstruction: the SP loss directly minimizes the square of the SP residual, whereas the KDK loss minimizes an MSE between the generated SR sample and its corresponding first-order KDK-evolved target. The latter yields smoother reconstructions, resulting in a modest suppression of power at the highest wavenumbers.

The fractional power-spectrum errors further highlight this difference. For the SP model, $(P_{\rm SR}/P_{\rm HR}-1)$ remains close to zero over most of the resolved range, indicating little systematic bias. The KDK model instead shows a more pronounced negative bias at intermediate and high wavenumbers, corresponding to an underprediction of small-scale power. Such behaviour is expected when the reconstruction favours smoother solutions and suppresses high-frequency fluctuations.

The SP model also consistently exhibits small power-spectrum errors across the shown test realizations. The KDK model shows slightly larger realization-to-realization variation, suggesting that the weaker physical constraints allow greater sensitivity to individual field realizations.

Overall, the models produce physically plausible high-resolution density fields and recover most of the missing small-scale information. However, the SP-trained model consistently provides a closer match to the HR targets in consideration, yielding sharper structures, improved recovery of high-$k$ power, and smaller systematic errors. These results indicate that incorporating the full SP dynamics improves the reconstruction of small-scale features compared to the first-order KDK approximation.

\paragraph{Power-spectrum statistics across test realizations}Figure \ref{fig7} shows the mean fractional power-spectrum error, ($P_{SR}/P_{HR}-1$), averaged over the test realizations, together with the corresponding 1$\sigma$ scatter. Both models reproduce the target power spectrum accurately over a broad range of wavenumbers, with mean errors remaining close to zero on large and intermediate scales. The scatter increases toward high-$k$, reflecting the increasing difficulty of reconstructing structures below the resolution limit of the LR input.

The SP-based model (left panel) exhibits a small positive bias at high wavenumbers, indicating a tendency to slightly overpredict small-scale power. This behaviour is accompanied by a larger realization-to-realization scatter, suggesting that the model reconstructs additional high-frequency structure whose amplitude varies between realizations. By comparison, the KDK-based model (right panel) remains closer to zero over most of the resolved range and shows a smaller systematic bias, although it develops a mild suppression of power at intermediate scales before rising at the highest $k$.

A notable difference is the dispersion of the error curves. The SP model displays a broader distribution at high wavenumbers, whereas the KDK model exhibits a more compact scatter. Unlike the representative samples shown in Figs.~\ref{fig5} and \ref{fig6}, which illustrate individual realizations where the SP model more closely matches the high-$k$ power spectrum, the ensemble statistics reveal that the SP model exhibits both a larger positive bias and greater realization-to-realization variance across the full test set. This behaviour is consistent with the optimization histories shown in Fig.~\ref{fig3}. The SP-residual loss is more challenging to optimize, resulting in a larger test physics loss throughout training. The increased optimization difficulty is reflected in the ensemble power-spectrum statistics through larger bias and variability in the reconstructed small-scale power. By comparison, the KDK loss provides a more stable optimization objective, yielding smoother reconstructions with smaller bias and reduced variance across the test set.

Overall, both models recover the target power spectrum with good accuracy up to $k \simeq 100 \, \textrm{h}\, \mathrm{Mpc}^{-1}$ . The SP model captures more small-scale power but at the cost of larger variance between realizations, whereas the KDK model yields slightly more stable predictions while tending to smooth the smallest resolved structures.

\section{Discussion and Future Directions}\label{Disc}
\subsection{Validity of Initial Conditions}
In most FDM simulations, initial conditions are generated exactly as described in this paper: first, the power spectrum is constructed using \textsc{AxionCAMB} for the given axion mass, then the initial conditions are usually obtained from the N-body particle distribution via density assignment such as the cloud-in-cell algorithm. The second step is especially convenient due to the availaibility of public codes (such as \textsc{JaxPM} for our case) for an arbitrary power spectrum.

However, this is not an ideal way to generate the initial conditions for field-based simulations. The reasons are twofold: first, the velocities of these N-body particles are typically computed using 2LPT method or Zeldovich approximation which are suited for CDM perturbations and secondly, converting from positions and velocities to fields incur some mesh error for FDM simulations. In this study, we nevertheless adopt this conventional particle-based initialization procedure because of its straightforward implementation and because we wanted to evaluate the performance of the ML models used here in regard to the inclusion of physics loss. In future work, we plan to generate more reliable initial conditions, following the methodology presented in the appendix of \citet{2025_Luu}. 

\subsection{Unsupervised Learning}\label{SS_sec}
The approach adopted in this work is semi-supervised, combining simulation-derived ground-truth data with physics-based constraints during training. A natural question is whether accurate cosmological simulations can be learned solely from the governing physical laws, without any supervised data. For relatively simple systems, such as gravitational evolution of a single Fourier-mode density perturbation, our previous work on SPINN demonstrated that fully physics-driven learning is feasible. However, extending this approach to realistic cosmological settings, characterized by complex initial conditions, an expanding background, and long evolutionary timescales, remains challenging. In these regimes, the optimization landscape becomes highly non-convex, and current optimization strategies are often insufficient for training fully unsupervised PINNs. An autoregressive framework that predicts the system evolution one step at a time using only physical constraints could be a promising alternative, though controlling the accumulation of errors over long time horizons would be a significant challenge.%, which can lead to significant deviations from the true physical evolution.

\subsection{Alternative super-resolution training}\label{Aux}
Our implementation of the super-resolution task relies on high-resolution (HR) FDM numerical simulations for the training objective. However, these simulations become prohibitively computationally expensive with increasing box sizes. An alternative is to train a generative model using only the physics loss (without any data loss), which leaves the solution space largely unconstrained and allows the network to generate multiple high-resolution realizations that satisfy the governing physics, rather than converging to a unique solution. Including a low-resolution data loss would  further constrain the admissible solution space by requiring the generated high-resolution fields, when downsampled, to agree with the corresponding low-resolution inputs. The primary challenge is that these two losses often define competing optimization objectives, resulting in conflicting gradient updates that can hinder convergence and lead to suboptimal solutions.

\section{Conclusion}\label{conc}
In this work, we present the first physics-informed generative framework for FDM simulations that explicitly incorporates the underlying physical dynamics during training. The framework addresses two complementary tasks: the evolution of cosmological fields from initial conditions to a target scale factor and the super-resolution of FDM simulations. Across both tasks, incorporating physics-informed losses consistently improves the physical fidelity of the generated fields.

For the evolution task, the model predicts the wavefunction and gravitational potential at a target redshift from realistic cosmological initial conditions in a $1\,h^{-1}\,\mathrm{Mpc}$ periodic cosmological box. We demonstrate that incorporating the SP physics loss substantially reduces the amount of training data required while accurately reproducing the reference simulations generated using a second-order KDK pseudo-spectral solver. The model remains stable throughout the trained redshift interval despite being trained on only 20\% of the available snapshots within the corresponding redshift range. An important advantage of the evolution model is that it eliminates the need to store the densely sampled intermediate snapshots of FDM simulations, which can require several terabytes of storage for large-scale simulations.

For the super-resolution task, we trained generative models using both the SP-residual and KDK-based physics losses. Both models accurately recover the target power spectrum up to $k \simeq 100\,h\,\mathrm{Mpc}^{-1}$, with mean power-spectrum errors below 20\% for most Fourier modes on previously unseen test realizations. While the SP model more faithfully preserves small-scale power and wave-interference structures, the KDK model produces smoother and more stable reconstructions at the expense of suppressing the smallest resolved scales. These results demonstrate that physics-informed training substantially improves the physical fidelity of generative super-resolution models for FDM.

Extending the present framework to larger simulation volumes and longer evolutionary times constitutes a natural next step. However, large-scale FDM simulations are unavailable because pseudo-spectral SP solvers require increasingly high spatial resolution to accurately resolve small-scale wave dynamics. While the inclusion of physics-informed losses appears to improve the quality of the reconstruction with limited high-resolution training data, further investigation is needed to determine whether this dependence can be eliminated entirely. Achieving this would enable low-resolution pseudo-spectral simulations to serve as the training data, with the generative model recovering the unresolved small-scale structure through the governing SP physics.

Looking beyond the current framework, neural operators provide a promising direction for further improving the scalability of physics-informed FDM emulators. In particular, autoregressive neural operator architectures, combined with appropriate physical constraints, may provide an efficient framework for learning the evolution of cosmological fields directly from the initial conditions while maintaining physical consistency. Future work will therefore investigate the integration of neural operators with physics-informed generative modeling, together with physically consistent cosmological initial conditions, to extend Cosmo-SPINN to substantially larger spatial domains and longer evolutionary times.

\section*{Acknowledgments}
The authors acknowledge financial support from the SNSF under the Starting Grant project Deep Waves (218396). The authors thank Philip Mocz for his valuable discussions during the project, especially for the generation of initial conditions. This work was supported by the use of the facilities of the Swiss National Supercomputing Centre (CSCS) under project sk029, and by the Swiss State Secretariat for Education, Research and Innovation (SERI) through the Swiss SKA Regional Centre (SKACH) Consortium.
\section*{Code Availability}
All models are implemented in PyTorch \citep{2019_Paszke}, and are trained
separately on a single NVIDIA Grace Hopper GH200 GPU. Code for this work (along with the scripts to produce the plots) is publicly available on Github at the following address: (\href{https://github.com/mishraashu6566/DeepWaves.Cosmo-SPINN.git}{DeepWaves.Cosmo-SPINN}).
\bibliography{mybib}{}

%%%%%%%%%%%%%%%%%%%%%%%%%%%%%%%%%%%%%%%%%%%%%%%%%%%%%%%%%%%%%%%%%
\appendix 
\section{Evolution model stability across redshifts}\label{appendix_A}

\begin{figure}
    \centering
    \includegraphics[width=\linewidth]{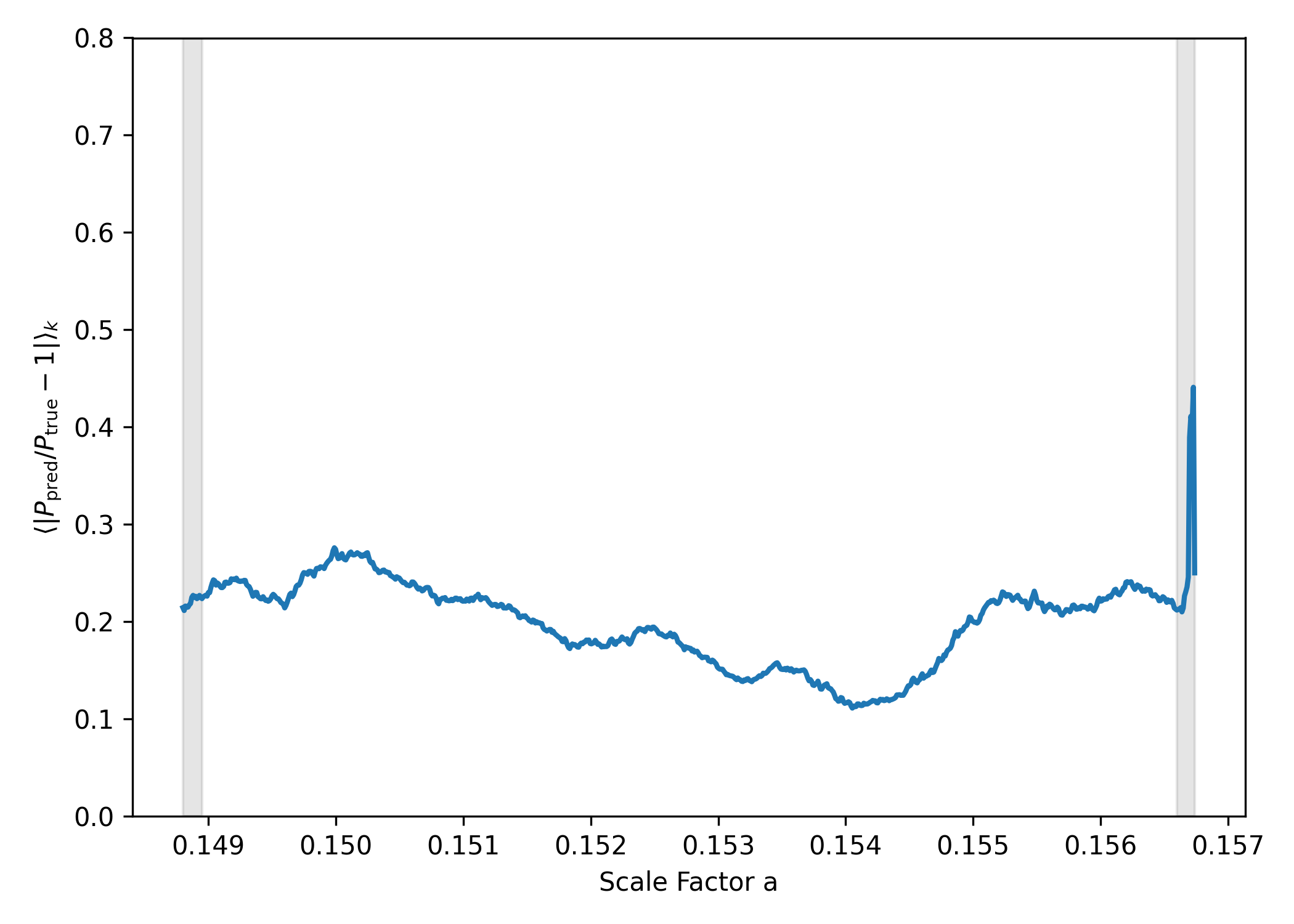}
    \caption{Mean absolute relative error in the predicted power spectrum, $\langle|\textrm{P}_\textrm{pred}/\textrm{P}_\textrm{true}-1|\rangle$ as a function of the scale factor a. The shaded vertical bands indicate the boundaries of the full data interval. The prediction error is minimized within the central region and increases toward the edges of the sampled range.}
    \label{fig8}
\end{figure}

Here we test the stability of evolution model across redshifts. This is particularly important because the training set contains only 20\% of the redshifts considered in the full dataset. The results are shown in Figure~(\ref{fig8}).

We find that the model remains stable over most of the sampled interval, but its performance deteriorates near the boundaries.
This behavior is expected. During training, redshift samples are randomly drawn between the minimum and maximum scale factors of the dataset. For points in the interior of the interval, training samples exist on both sides, allowing the network to effectively interpolate. In contrast, boundary points have neighboring samples only on one side, making interpolation more challenging. In addition, random sampling naturally results in denser coverage of the interior than the edges of the interval. Furthermore, the finite-difference approximation used to evaluate derivatives near the boundaries effectively pushes the problem into a mild extrapolation regime, where the PDE residuals alone do not provide sufficiently strong constraints. These effects together lead to the increased prediction errors observed at the boundaries.

\section{Generalization of the Evolution Model to Unseen Initial Conditions}\label{appendix_B}
We extended the evolution model, where the network takes the initial condition and outputs the wavefunction and the gravitational potential at the specified scale factor, to multiple realizations of the initial conditions just as we did for the super-resolution task. This constitutes a more challenging learning problem than super-resolution, since the network does not receive a low-resolution approximation of the target field. Instead, it must learn the full dynamical evolution from the initial conditions and generalize this mapping across unseen realizations.

The training and test data are split in an exactly same way as described for the super-resolution case, but with 100 realizations in total instead of 80. 
The architecture has been kept the same, and the network outputs the residual $\delta$ where:
\begin{equation}
    \hat{\psi}(a) = \psi(a_0) + \,\delta,
\end{equation}
The factor of 0.1 previously included in Eq.~\eqref{Ev_1} was removed, as it produced predicted fields with excess power relative to the reference fields. Empirically, removing this scale factor yielded better performance while also proving effective for the super-resolution task. We keep the rest of the neural network architecture, optimizer, and all hyperparameters identical to the setup used for the evolution model used for single realization training.

\begin{figure*}
    \centering
    \includegraphics[width=\linewidth]{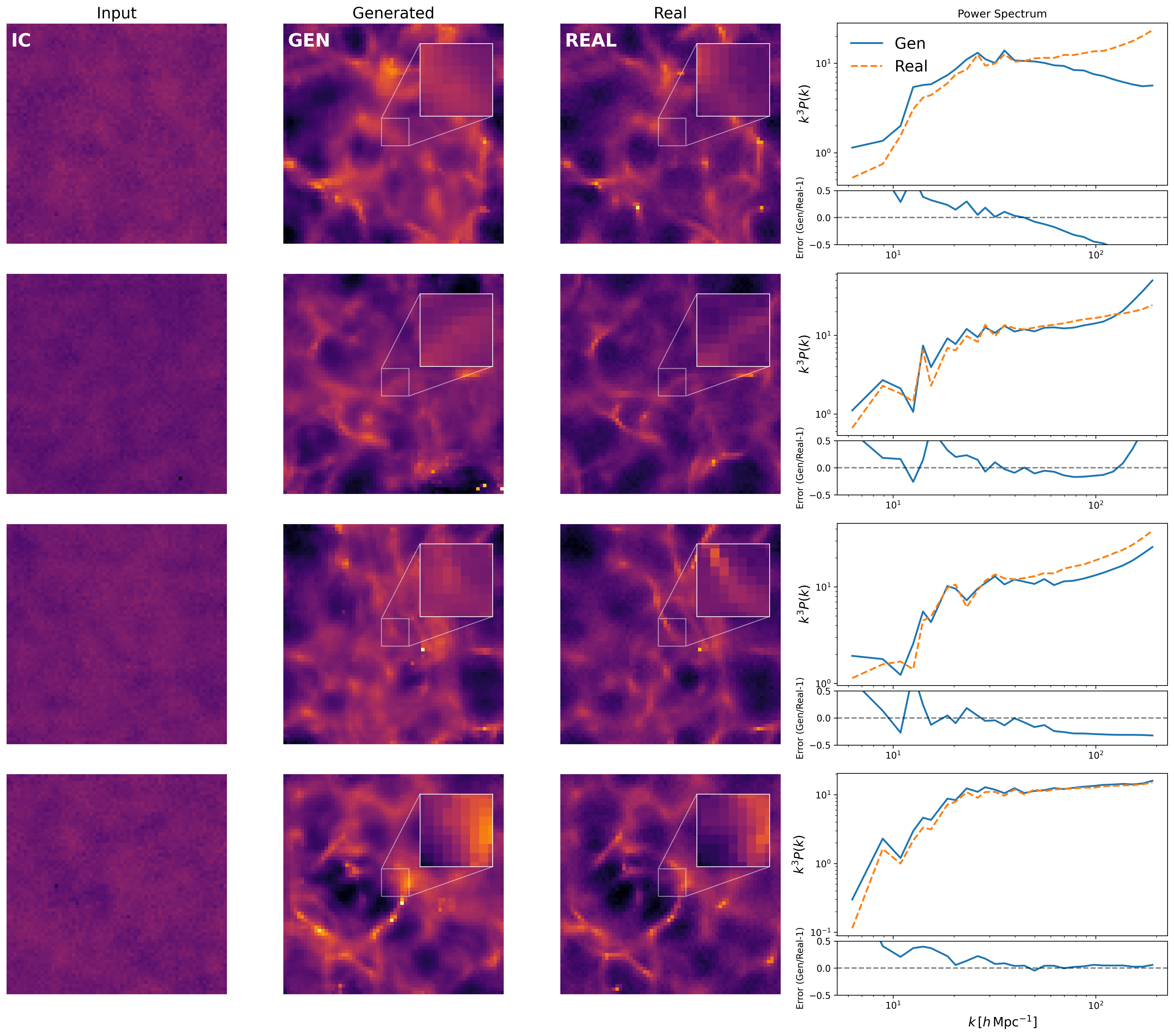}
    \caption{Examples of density fields generated by evolution model \textit{the unseen test realizations}. The  generated  fields ($64^3$, Gen), and target density fields ($64^3$, Real) are shown as density projections along the z-axis on a logarithmic scale. Insets show zoomed-in regions. The right panels compare the dimensionless power spectra $k^3P(k)$ of the Gen, and Real fields, while the lower panels show the fractional error ($P_{Gen}/P_{Real}-1$).  The model recovers roughly the target power spectrum over a wide range of scales, except with some discrepancy at small scales}
    \label{fig9}
\end{figure*}

Fig.(\ref{fig9}) shows the generated fields and the and the corresponding dimensionless power spectra ($\Delta(k)=k^3P(k))$ predicted by the evolution model trained  using the SP physics loss. The spectra are plotted up to the Nyquist frequency of the grid ($\simeq 402 \, \textrm{h}\, \mathrm{Mpc}^{-1}$). As before, the fundamental mode is $(k_f \simeq 6 \, \textrm{h}\, \mathrm{Mpc}^{-1})$, implying that approximately 60\% of the resolved modes ($k_f < k < k_J$) lie in the gravity-dominated regime. We see that even in this case of generalization of initial conditions, the model can correctly recover the large scales of the target fields. However, discrepancies persist on small scales, primarily because learning the wavefunction phase is particularly challenging. The physics-informed loss alone admits multiple solutions that satisfy the governing equations but differ in phase from the training data, allowing the network to converge to physically consistent yet phase-shifted solutions.

\end{document}